\documentclass[iop,apj]{emulateapj}
\usepackage{graphicx,natbib}
\usepackage{longtable}

\begin{document}

\title{Wave propagation and jet formation in the chromosphere}

\author{L. Heggland}
\affil{Institute of Theoretical Astrophysics, University of Oslo,
  P.O. Box 1029, Blindern, N-0315 Oslo, Norway}
\email{lars.heggland@astro.uio.no}

\author{V. H. Hansteen\altaffilmark{1}}
\affil{Institute of Theoretical Astrophysics, University of Oslo,
  P.O. Box 1029, Blindern, N-0315 Oslo, Norway}

\author{B. De Pontieu}
\affil{Lockheed Martin Solar and Astrophysics Laboratory, 3251 Hanover Street,
  Org. ADBS, Building 252, Palo Alto, CA 94304}

\and

\author{M. Carlsson\altaffilmark{1}}
\affil{Institute of Theoretical Astrophysics, University of Oslo,
  P.O. Box 1029, Blindern, N-0315 Oslo, Norway}

\altaffiltext{1}{Also at Center of Mathematics for Applications,
  University of Oslo, P.O. Box 1053, Blindern, N-0316 Oslo, Norway}

\begin{abstract}

We present the results of numerical simulations of wave propagation
and jet formation in solar atmosphere models with different magnetic
field configurations. The presence in the chromosphere of waves with
periods longer than the acoustic cutoff period has been ascribed to
either strong inclined magnetic fields, or changes in the radiative
relaxation time. Our simulations include a sophisticated treatment of
radiative losses, as well as fields with different strengths and inclinations.
Using Fourier and wavelet analysis techniques,
we investigate the periodicity of the waves that travel through
the chromosphere. We find that the velocity signal is dominated by
waves with periods around 5~minutes in regions of strong, inclined field,
including at the edges of strong flux tubes where the field expands,
whereas 3-minute waves dominate in regions of weak or vertically
oriented fields. Our results show that the field inclination is very
important for long-period wave propagation, whereas variations in the radiative
relaxation time have little effect. Furthermore, we find that atmospheric
conditions can vary significantly on timescales of a few minutes, meaning
that a Fourier analysis of wave propagation can be misleading. Wavelet
techniques take variations with time into account and are more suitable
analysis tools. Finally, we investigate the properties of jets formed
by the propagating waves once they reach the transition region, and find
systematic differences between the jets in inclined field regions and those in
vertical field regions, in agreement with observations of dynamic
fibrils.

\end{abstract}


\section{Introduction}

Chromospheric wave propagation has been an extensively studied, but poorly
understood, subject in solar physics for some time. One puzzle has been
the presence of propagating waves with periods on the order of 5 minutes
or more. Such waves were observed by, e.g., \citet{Orrall1966} and
\citet{Giovanelli+etal1978};
this was considered surprising since the acoustic cutoff period in the
chromosphere, above which waves should not be able to propagate, is on the
order of 200~s. Such long-period propagation was later found to be widespread,
generally occurring wherever the local magnetic field is strong.
\citet{Lites+etal1993} show an example of observations of neighboring
internetwork and network regions that have a marked difference in their
Fourier spectra.

It was realized \citep{Michalitsanos1973,Bel+Leroy1977,Suematsu1990} that,
since magnetoacoustic waves in a strongly magnetized medium are restricted
to propagating along field lines, the effective gravity (i.e., the component of
gravity along the magnetic field) would be reduced in magnetic regions. Since
the cutoff frequency (in an isothermal atmosphere) is given by
\begin{equation}
  \nu_{ac}=\frac{\gamma g_{eff}}{4\pi c_s},
  \label{cutoffeq}
\end{equation} where
$c_s$ is the sound speed and $\gamma$ is the ratio of specific heats, the
cutoff frequency would also be lower in regions of strong inclined field,
potentially allowing 5-minute ($\sim$3~mHz) waves to propagate.
This hypothesis has since
been tested in a number of increasingly advanced numerical simulations, e.g.
\citet{DePontieu+etal2004}, \citet{Hansteen+etal2006}, and
\citet{Heggland+etal2007}. All have found that it is an effective mechanism
for transmitting 5-minute power through the chromosphere; some models
have suggested that the leakage of 5-minute waves into the chromosphere
can also explain the presence of 5-minute oscillatory signal in the corona
\citep{DePontieu+etal2005}.

One criticism of this explanation has been that not all observations of 5-minute
propagation are in regions of obviously inclined field.
An alternative explanation has therefore been suggested,
originally by \citet{Roberts1983} and later developed and tested by
\citet{Centeno+etal2006,Centeno+etal2009} and \citet{Khomenko+etal2008},
in which changes in the radiative relaxation time associated with small scale
magnetic structures are responsible for increasing the cutoff
period. This mechanism has also been demonstrated to enable propagation of
5-minute oscillations, even in vertical magnetic structures. However, the
underlying basis of this theory and the related simulations is a highly
simplified energy equation in which radiative
losses are approximated by Newton's law of cooling
(\citeauthor{Khomenko+etal2008} \citeyear{Khomenko+etal2008}
and references therein).

A related subject that has been studied extensively is the generation and
propagation of chromospheric jets such as spicules (type I and II),
macrospicules, surges, fibrils, and mottles. A variety of models have
been proposed over
the years (see \citeauthor{Sterling2000} \citeyear{Sterling2000} for a review), and although it remains
unclear whether the many types of jets are different manifestations
of the same underlying physical phenomenon or not, there are at least
strong indications that the jets known as dynamic fibrils are driven
by shock waves traveling through the chromosphere
\citep{Hansteen+etal2006,DePontieu+etal2007,Heggland+etal2007,Martinez+etal2009a}.
Many of these jets have lifetimes around 5~minutes, and so the waves
driving them need to have been channeled into the upper chromosphere
via one of the processes outlined above.

In this paper, we present the results of two-dimensional simulations
of wave propagation from the convection zone to the transition region
and corona. The simulations include a sophisticated treatment of radiative
losses and heat conduction, and study the effects of different magnetic
field geometries and strengths on the propagation of waves through
the chromosphere. We also look at the jets that these waves produce
once they reach the transition region, and perform a statistical
comparison of the jets produced in a model with vertical field and in
one with inclined field.

In \textsection{}2 we describe the different simulations and the code.
\textsection{}3 contains an analysis of wave propagation and periodicities
in the various models, and the results are discussed in \textsection{}4.
\textsection{}5 looks at jet formation and properties, and a summary
and conclusions follow in \textsection{}6.

\section{Simulations}

The simulations have been run using the {\it Bifrost} code;
a detailed code description has recently been published
\citep{Gudiksen+etal2011}. It is a 3D MHD code that is designed
to model the solar atmosphere to a high degree of realism by
including as many of the relevant physical processes as practical and
possible. The code is staggered such that scalars are defined on the
grid interior while fluxes are defined on the edges between
computational cells. Sixth order polynomials are used to calculate derivatives.
High order interpolation is also used when a
given variable needs to be evaluated half a grid cell away. In order
to maintain stability, artificial viscosity and resistivity terms with
strengths that can be set by the user are included following a
hyperdiffusive scheme, such that regions of large gradients are also
those with the largest viscosity and magnetic diffusivity. This
allows the diffusivity to be low throughout most of the simulation
box, only becoming significant where the conditions require it.

\begin{deluxetable}{lccl}
\tablecolumns{4}
\tablewidth{0pt}
\tablecaption{Summary of simulation properties \label{simtab}}
\tablehead{
  \colhead{Model} & \colhead{Grid cells} & \colhead{Dimensions [Mm]} & \colhead{Magnetic field}}
\startdata
Case A & $512 \times 325$ & $16.61 \times 15.80$ & Extremely weak \\
Case B & $512 \times 325$ & $16.61 \times 15.80$ & Moderate, vertical \\
Case C & $400 \times 535$ & $11.17 \times 14.08$ & Strong expanding tube \\
Case D & $512 \times 325$ & $16.61 \times 15.80$ & Moderate, inclined \\
Case E & $512 \times 325$ & $16.61 \times 15.80$ & Strong, inclined
\enddata
\end{deluxetable}

The code includes thermal conduction along magnetic field lines, computed
implicitly using a multigrid algorithm. A ''realistic'' equation of
state is pre-computed based on LTE, and stored in tabular form to
compute temperatures, pressures, and radiation quantities, given the
mass density and internal energy. The code solves the equations of
radiative transfer for the photosphere and lower chromosphere,
including scattering, using the multigroup opacity methods developed
by \citet{Nordlund1982} and \citet{Skartlien2000} in short characteristic form
\citep{Hayek+etal2010}. In the upper chromosphere, transition region, and
corona, we utilize an advanced radiative loss function consisting of several
parts. In the corona and transition region, the optically
thin approximation is used, taking into account radiation from
hydrogen, helium, carbon, oxygen, neon, and iron using atomic collisional
excitation rates from the HAO-DIAPER atom data package
\citep{Judge+Meisner1994}. In the upper chromosphere, most important
lines are not in local equilibrium, and excitation rates are
precomputed and tabulated from 1D chromospheric models with radiative
transfer
calculated in detail \citep[e.g.,][]{Carlsson+Stein1995,Carlsson+Stein1997}.
More details on these radiative loss functions as well as on the code
itself can be found in \citet{Leenaarts+etal2011} and \citet{Gudiksen+etal2011}.

In sum, the code thus includes most~physics~that~are important in the 
solar atmosphere; the main omission in the present work is time-dependent
ionization, which mainly has an effect in the upper chromosphere and transition region, above a height of $\sim$1~Mm. See \citet{Leenaarts+etal2007,Leenaarts+etal2011}
for a discussion of these effects.

The horizontal boundary conditions are periodic, while both the vertical
boundaries are open. The upper one is based on characteristic
extrapolation of the magnetohydrodynamic variables; the lower one allows
outflowing material to leave the box, while the entropy of the inflowing
material is set.
There is no piston or other imposed external driver producing wave motions;
all oscillations are produced self-consistently by the turbulent motion
in the convection zone.

\begin{figure}
  \includegraphics[scale=0.60]{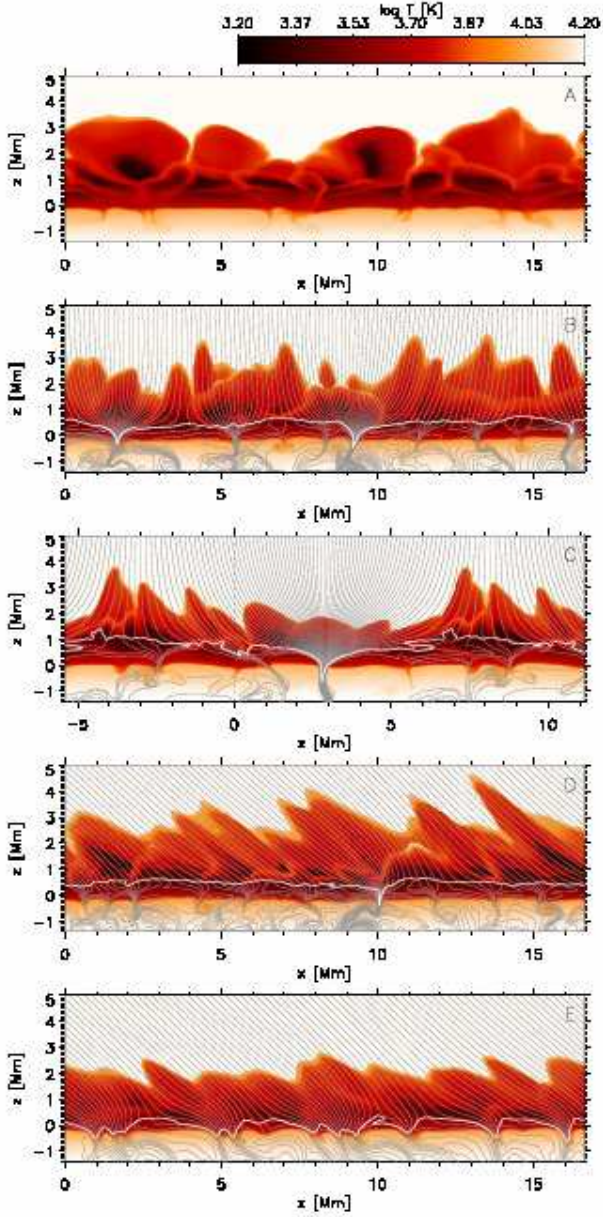}
  \caption{Temperature plot showing the initial states of our five
           different simulations.
           The white lines are at the height where the plasma $\beta=1$,
           while the gray lines are magnetic field lines. Case A is
           high-$\beta$ throughout and contains only a very weak
           magnetic field. The dotted vertical black line in case
           C marks the location of the horizontal boundary; as this
           model has a smaller horizontal extent than the others, the
           rightmost 5.5~Mm have been repeated on the left.}
  \label{initfig}
\end{figure}

\begin{figure}
\begin{center}
  \includegraphics[scale=0.8]{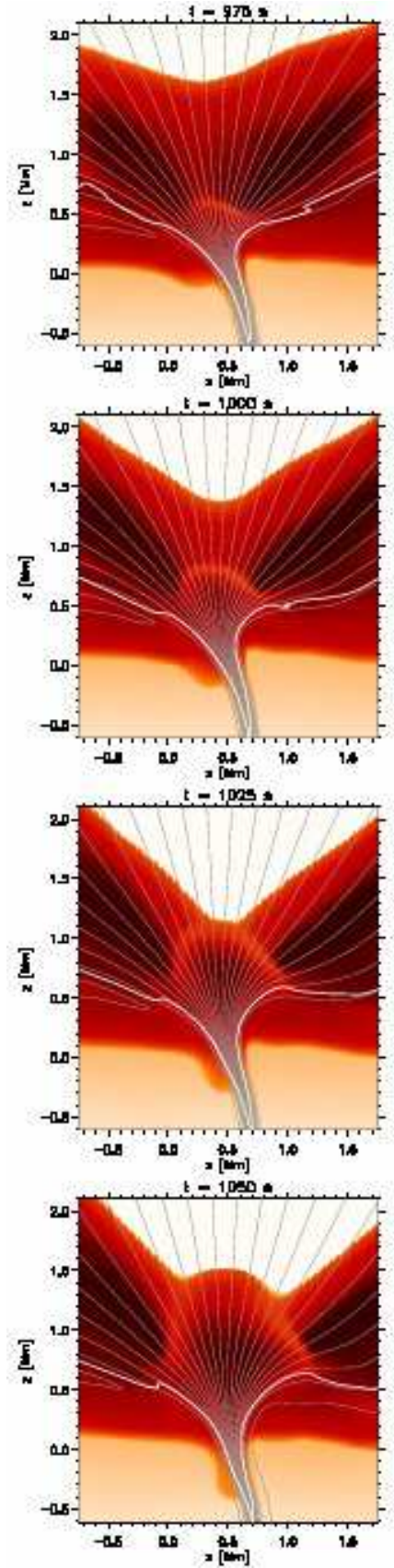}
  \caption{Temperature plot showing a wave propagating upwards and
           outwards in an expanding
           flux tube in case C. The white line is at the height where
           the plasma $\beta=1$, while the gray lines are magnetic
           field lines.}
  \label{propagation}
\end{center}
\end{figure}

\begin{figure*}
  \plotone{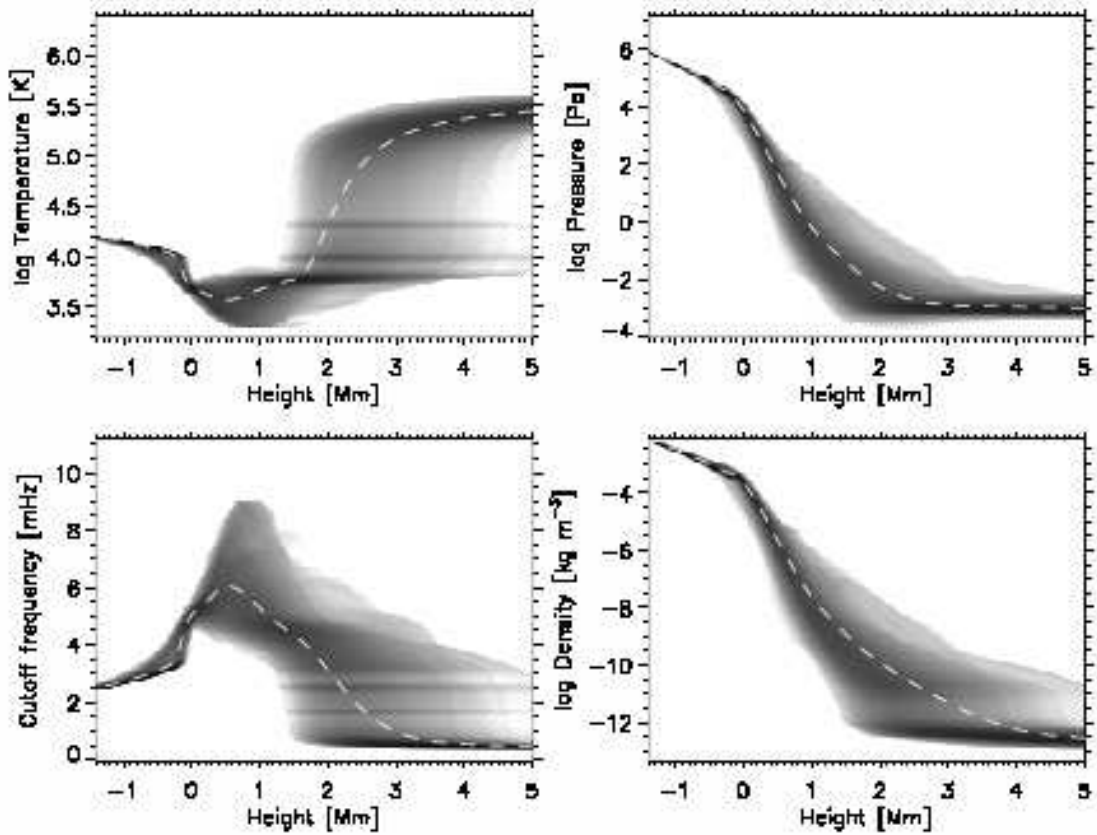}
  \caption{Temperature (top left), cutoff frequency (bottom left), pressure
           (top right), and density (bottom right) in case D.
           These are scatter
           plots showing how often the atmosphere is in a certain state
           (across all times and horizontal positions);
           more densely populated states are darker. The white
           dashed lines are the averages over all times and horizontal
           positions.}
  \label{tprc}
\end{figure*}

\begin{figure}
  \plotone{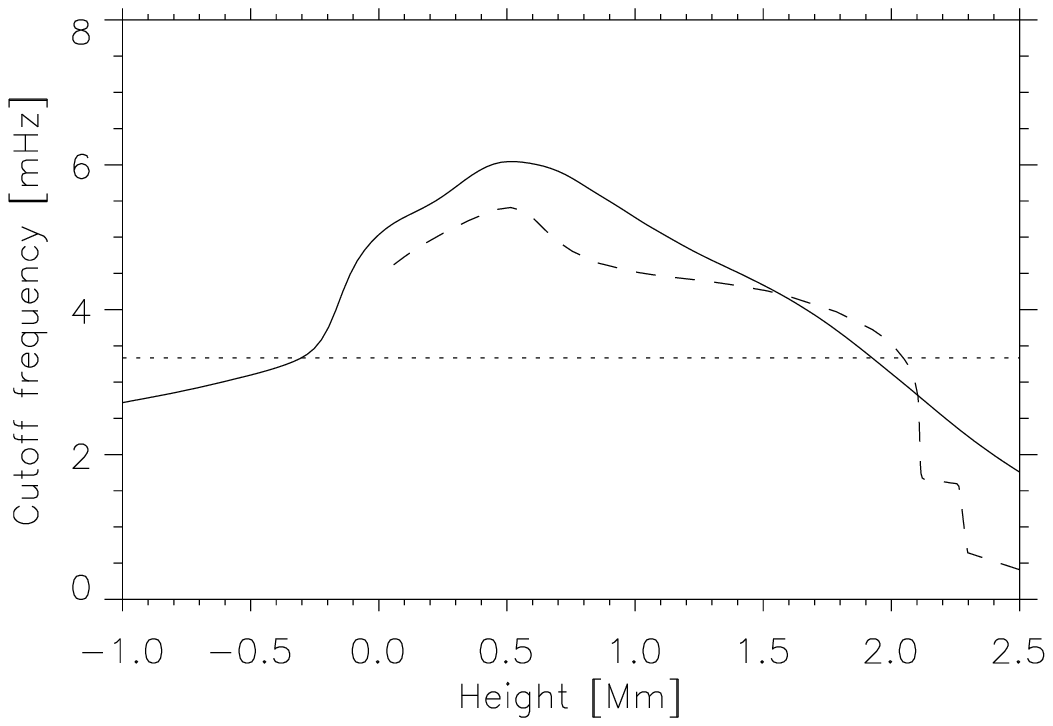}
  \caption{Average cutoff frequency in our case D (solid line) and in the
           VAL3C solar atmosphere model (dashed line). The horizontal dotted
           line is at $3.33$~mHz, corresponding to a period of 5 minutes.}
  \label{cutoffval}
\end{figure}
 
We will be analyzing five different simulations, using model
atmospheres that are about 16~Mm high. They extend
from the upper layers
of the convection zone at the lower boundary ($z=-1.4$~Mm), through the 
photosphere ($z=0$~Mm corresponding to the average height where
$\tau_{5000}=1$), chromosphere, transition region and corona, with the upper
boundary at $z=14.4$~Mm. The simulation boxes are two-dimensional and 
contain $512 \times 325$
grid cells, using a uniform horizontal spacing of 32.5~km.
In the vertical direction, we use a non-uniform spacing which is 28~km
between each cell from the
lower boundary to $z=5$~Mm, increasing gradually until
$z=9.6$~Mm; from there to the upper boundary it stays at a constant 
value of 150~km.
The properties of our five different simulations are summarized in
Table~\ref{simtab}, and Figure~\ref{initfig} shows their initial temperature
and magnetic field structures. Case C has slightly smaller dimensions and 
higher resolution than the others; the other models differ
mainly in the strengths and orientations of their magnetic fields.
All the simulations have been run for 4500~s, or 75 minutes, of solar time
(after an initial relaxation period).

The turbulent motion in the convection zone is able to move magnetic flux
tubes around, which means that the field configurations in the high-$\beta$
sections (where $\beta$ is the ratio between the gas and magnetic pressures)
of all five simulations are rather dynamic. In contrast, the field
in the upper chromosphere and corona in all models except case A
is rather stable and changes little with time. The motion in the convection
zone also generates waves that propagate upwards (see
Figure~\ref{propagation}) and create jets of various
lengths and lifetimes once they reach the transition region. Many such
jets can be seen in Figure~\ref{initfig}. They tend to follow the magnetic
field, taking on relatively thin, elongated shapes, but in the non-magnetic
case A, they look more like extended spherical waves moving the transition
region up and down.

Since we are primarily interested in wave propagation in the
chromosphere, not the corona, we omit from our analysis the regions above
$z=5-6$~Mm, the height of the peaks of the tallest jets.

In Figure~\ref{tprc} we show the
temperature, pressure, and density structures of case D in greater detail
as functions of height, as well as the cutoff frequency (given by
equation~\ref{cutoffeq}). The other cases look similar.
The temperature plot clearly shows the
significant movement the transition region undergoes as jets are continually
formed and push it upwards. The three prominent horizontal lines in the
temperature structure correspond to the ionization temperatures of
hydrogen and helium (the latter twice), and so are an effect of the
tabulated equation of state \citep{Leenaarts+etal2011}. The plot of
the cutoff frequency
shows the region, on average between about $z=0$~Mm and $z=1.8$~Mm but
with large variation in time and space, where
the cutoff is above $3.33$~mHz, corresponding to a period of 5 minutes.
This is the region where such long-period waves are evanescent, and which
they will have to ``tunnel'' through if they are to survive into the upper
layers of the atmosphere. In Figure~\ref{cutoffval} we have plotted the
average cutoff frequency in our case D and the cutoff calculated from the
semi-empirical VAL3C model \citep{Vernazza+etal1981}. The VAL model
has a somewhat flatter cutoff in the high chromosphere than our average
model, but agrees that the cutoff frequency remains high until one
reaches the transition region at about $z=2$~Mm.

Equation~\ref{cutoffeq}, which we used in calculating the cutoff,
is strictly speaking
only valid in an isothermal atmosphere. There are many possible
definitions of the cutoff frequency
in a stratified atmosphere
\citep{Schmitz+Fleck1998,Erdelyi+etal2007}, and the values
in Figures~\ref{tprc} and \ref{cutoffval} should not be considered
exact. The main point is to illustrate that the solar atmosphere
contains a height range where the cutoff frequency is higher than the
frequencies associated with photospheric oscillations such as $p$-modes.

We should point out that such long-period waves have very long wavelengths,
so ``local'' conditions are not a very well defined term when analyzing them.
A wave with a 5-minute period will have a wavelength of 2100-3000~km,
given typical sound speeds in the lower atmosphere of 7-10~km~s$^{-1}$.
The entire height of the photosphere and chromosphere therefore amounts
to less than one wavelength. From this, we may expect that even in the
absence of any channeling, through field inclination or other processes,
a portion of the 5-minute power could make it through to the upper layers. 
Another expectation
is that such channeling does not necessarily have to be active throughout
the entire 2~Mm high region, though the effect will likely be greater the
more prevalent conditions conducive to channeling are (in both space and time).

This paper aims to bring greater clarity to issues such as the effectiveness
of channeling mechanisms and the effects of propagating long-period waves on
higher atmospheric layers. To that end,
our analysis focuses on two separate but related phenomena: first, the
propagation of waves of different periodicities through the atmosphere
and the influence upon them by the magnetic field, and second, the jets
produced by these waves once they reach the transition region. To study
the former, we perform an extensive Fourier and wavelet
analysis of the five simulations.
For the latter, we individually measure properties such as the lifetimes,
lengths and maximum velocities of the jets, and make a statistical 
comparison between
case B (vertical field) and case D (inclined field).

\section{Wave propagation}

\subsection{Case A}

\begin{figure}
\includegraphics[scale=0.43]{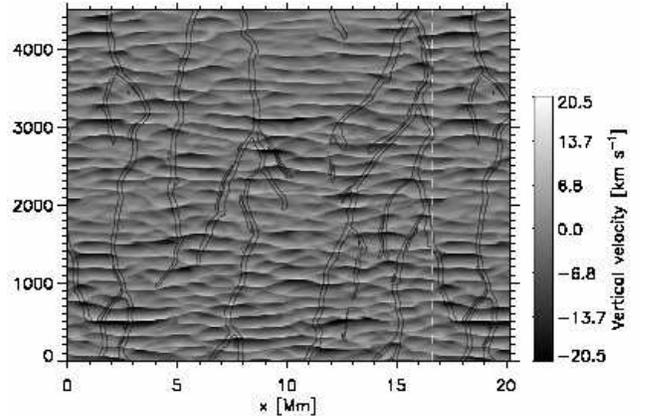}
  \caption{Vertical velocity at $z=1$~Mm in case A, with
    superimposed contours showing the regions where $B_z$ at
    $z=0$~Mm is greater than $0.3$~G. The white dashed line shows
    the location of the periodic boundary; the leftmost $3.5$~Mm
    are repeated on the right.}
  \label{weakBuz}
\end{figure}

\begin{figure}
  \plotone{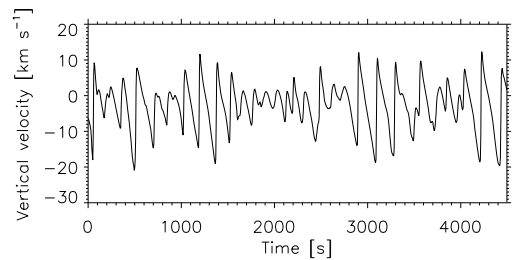}
  \caption{Vertical velocity at $z=1$~Mm and $x=1.5$~Mm in case A.}
  \label{weakBuz2}
\end{figure}

\begin{figure*}
  \plotone{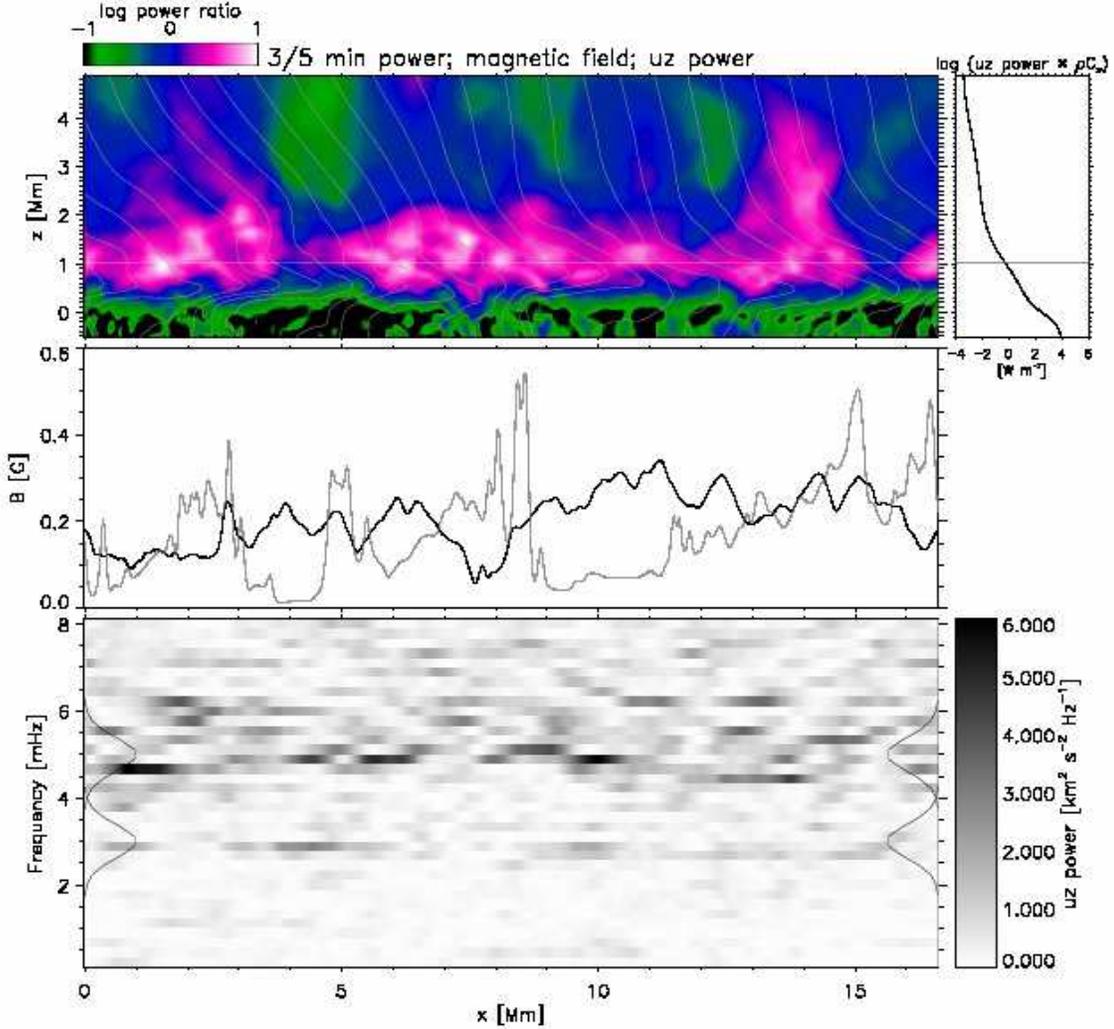}
  \caption{Fourier analysis of case A, showing (top) the ratio of 3-minute
    power to 5-minute power, with time-averaged
    magnetic field lines (thin gray lines) superimposed, as well as a
    horizontal line at $z=1$~Mm; (top right) the total energy flux as
    a function of height;
    (center) the time-averaged magnetic field strength at $z=1$~Mm (black)
    and $z=0$~Mm (gray); and (bottom) the power spectrum of the
    vertical velocity, also showing the Gaussian 3-minute (5~mHz)
    and 5-minute (3~mHz) bands.}
  \label{cuzpow}
\end{figure*}

We start by analyzing case A, a reference model that contains only
a very weak magnetic field. It is everywhere
weaker than 1~G, and the model's behavior is expected
to be essentially non-magnetic. For consistency with the other models,
however, the magnetic field is included in the
analysis below. The initial temperature state is shown in the top
panel of Figure~\ref{initfig}. This model is meant to be representative
of conditions in the internetwork.

The turbulent motion of the convection zone generates waves and flows
at many different periods and in many locations. Figure~\ref{weakBuz} shows
a plot of the vertical velocity\footnote{Upward velocity is taken to be
positive.} (grayscale) as a function of horizontal
position ($x$) and time, at a constant height ($z$) of 1~Mm. The
overplotted black contours, for consistency with the other cases,
mark the locations where the magnetic field is strongest; specifically,
where the vertical component of the magnetic
field at $z=0$~Mm (1~Mm below the height at which the velocity is plotted)
is $0.3$~G or greater. Obviously, these ``flux concentrations''
are still very weak. Since one of the flux concentrations
lies right on the (periodic) boundary, the plot shows the leftmost 3.5~Mm
of the box again on the right; the location of the boundary is marked with
a vertical white dashed line. In this case, the wavefronts appear to be
rather uniformly distributed, with no strong clustering into specific
regions either in time or space. They are typically horizontally
coherent on length scales of 3-5~Mm.

Figure~\ref{weakBuz2} shows the vertical velocity at the horizontal
position $x=1.5$~Mm, allowing us to see the wave shapes in more
detail. The waves are notably asymmetric, showing the characteristic
``N''-shape of compressive waves that have steepened into shocks as
a result of the density stratification. The amplitude varies quite
a bit with time due to the randomness of the photospheric driver.
The strongest shock trains appear around $t=1100$~s, $t=2900$~s, and
towards the end of the simulation. At these times, the shocks
reach peak-to-peak amplitudes of up to 30~km~s$^{-1}$, generally with stronger
downflows than upflows. In fact, the time-averaged velocity
at this location is a downflow of about 3~km~s$^{-1}$, although
this does not imply a net downward mass flux, since the shock
waves propagating upward have greater densities. The time
between velocity peaks is typically around 200~s.

We have created power spectra of the simulations by performing Fourier
transforms of the velocity at each grid cell. An example of the resulting
spectra is shown in Figure~\ref{cuzpow}. The lower panel shows the power
spectrum itself as a function of $x$ and frequency $\nu$ at our analysis height
of $z=1$~Mm, the same height where we plotted the
velocity field in Figures~\ref{weakBuz} and \ref{weakBuz2}. This height is also
marked by the light gray horizontal line in the upper panels.
The bell-shaped black curves on either side delineate the two dominant
frequency bands used in the subsequent analysis, the
``5-minute'' band centered at $3$~mHz and the ``3-minute'' band
centered at $5$~mHz. Both are Gaussian bands including contributions
at frequencies up to 1~mHz above and below the central frequency. The central
frequencies correspond to actual periods of 333~s and 200~s, respectively.

The upper panel shows the ratio between the power in the 3-minute
band and that in the 5-minute band, showing 3-minute dominance as pink and
white, 5-minute dominance as green and black, and approximately equal power
as blue. Superimposed on this plot are the height where the power spectrum 
in the lower panel is plotted
(light gray horizontal line), and magnetic field lines calculated from the
time-averaged field (dark gray). The small panel to the right shows the
height profile of the total velocity Fourier power across all frequencies,
multiplied by the $x$-averaged
density and sound speed to create a measure of the energy flux density
rather than the straight velocity power.
The middle panel of the figure shows the total time-averaged magnetic field
at $z=1$~Mm (black) and at $z=0$~Mm (gray).

In this model, the 3-minute band is dominant throughout
the chromosphere ($z=0.5-1.5$~Mm). Only in a small area near $x=15.5$~Mm
does the 5-minute band achieve rough parity, but a look at the spectrum
(lower panel) shows that the dominant frequencies at that point are at
or above 6~mHz, and begin to fall outside the 3-minute band. In fact,
there is significant power at frequencies of $6-7$~mHz across most of
the simulation box.

This picture, with 3-minute domination throughout, is typical of the
classical results from observations of weakly magnetic regions in the
internetwork (e.g., \citeauthor{Lites+etal1993} \citeyear{Lites+etal1993}
and references therein). It is
a natural consequence of the influence of the acoustic cutoff frequency and
the radiative damping of higher-frequency waves.
Under chromospheric conditions, the acoustic cutoff frequency
is typically around $4.5$-5~mHz, corresponding to periods of 200-220~s (see
Figure~\ref{cutoffval}), and
waves at lower frequencies will have difficulty propagating upwards.
It is, however, possible to see small amounts of 5-minute power 
across much of the box, particularly around $x=4$~Mm, $x=10$~Mm, and
$x=15$~Mm. This shows that although the long-period waves are damped,
their long wavelength still ensures that part of their power reaches
$z=1$~Mm.


\subsection{Case B}

\begin{figure}
\includegraphics[scale=0.43]{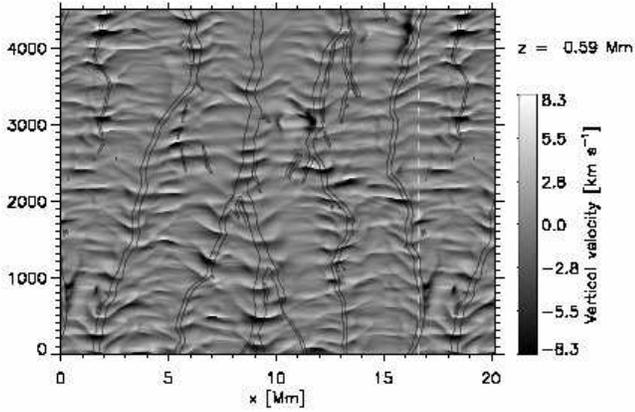}
  \caption{Vertical velocity at $z=0.59$~Mm in case B, with
    superimposed contours showing the regions where $B_z$ at
    $z=0$~Mm is greater than 300~G. The white dashed line shows
    the location of the periodic boundary; the leftmost $3.5$~Mm
    are repeated on the right. (An mpeg animation of this figure
    is available in the online journal.)}
  \label{vertBuz}
\end{figure}

Having studied an essentially non-magnetic model, we now turn our
attention to models with magnetic fields of varying strengths and
orientations. Case B is a model with several flux concentrations in
the photosphere. These flux concentrations are relatively narrow
(a few hundred km) and have strengths of $500-1000$~G.
Such concentrations are a result of the convective motions below
the photosphere, which tend to concentrate the magnetic field in
relatively narrow regions, often connected with an average downflow,
such as intergranular lanes. As we move upwards, the field expands,
creating inclined field lines on the edges of the flux concentrations
in the chromosphere, while once we reach coronal heights, the field
fills all space and the orientation is close to vertical. These effects
can be seen in the second panel from the top of Figure~\ref{initfig},
which shows the initial state of this case. This
model is intended to represent solar conditions in and around network
or plage regions.

\begin{figure}
\includegraphics[scale=0.43]{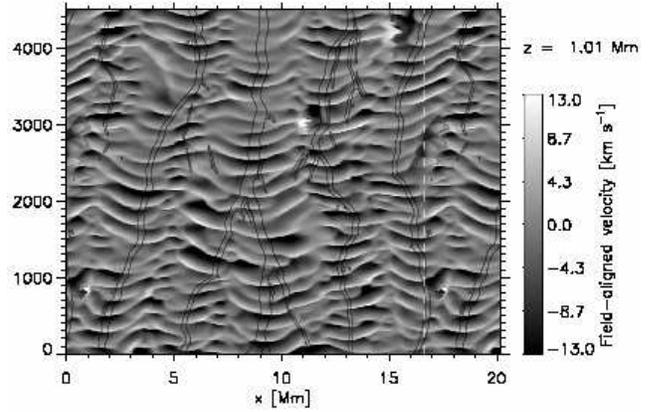}
  \caption{As Figure~\ref{vertBuz}, but showing the field-aligned
    (rather than vertical) velocity at $z=1$~Mm. This is the velocity
    that is used in the subsequent Fourier and wavelet analysis.}
  \label{vertBud}
\end{figure}

\begin{figure*}
  \plotone{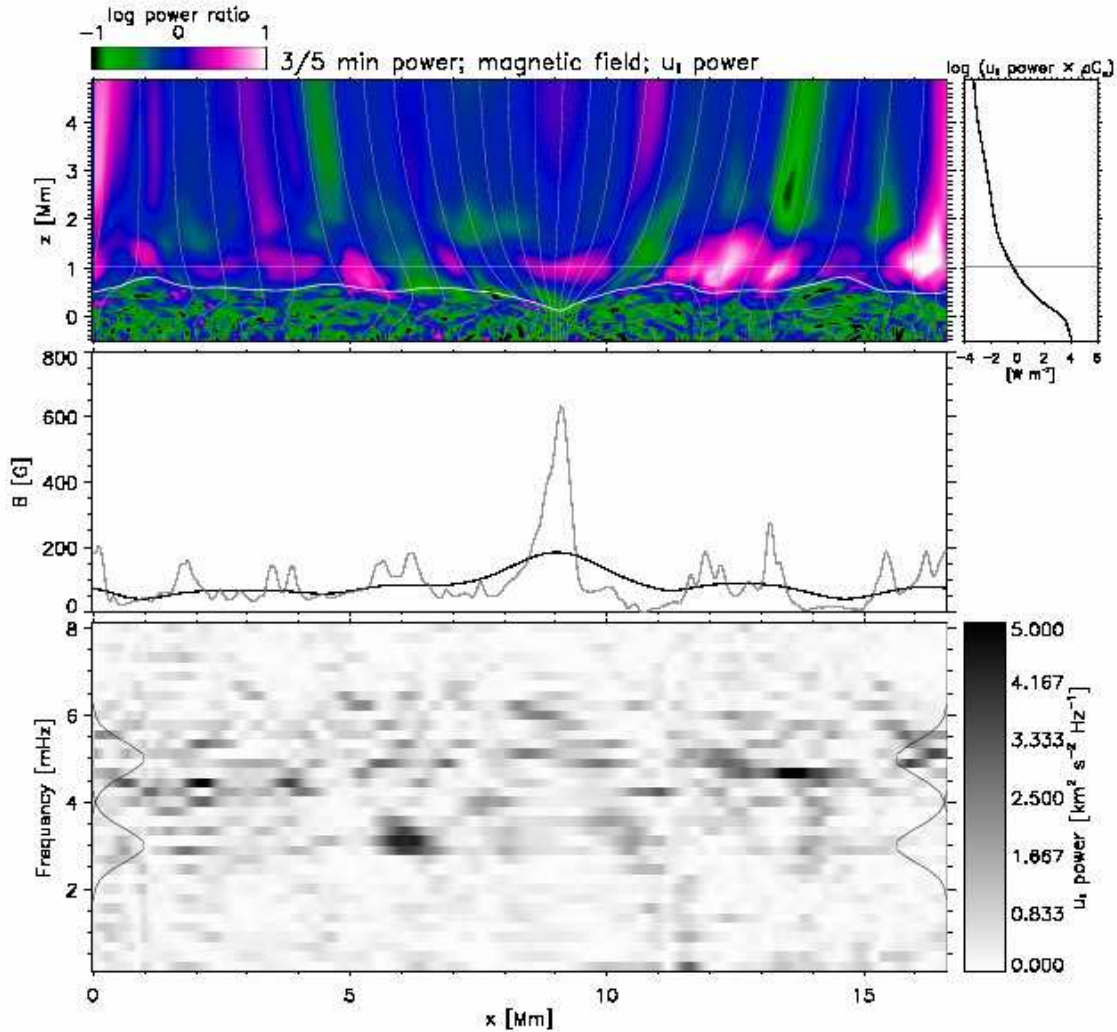}
  \caption{Fourier analysis of case B, showing (top) the ratio of 3-minute
    power to 5-minute power, with time-averaged $\beta$=1 surface (thick
    white line) and
    magnetic field lines (thin gray lines) superimposed, as well as a
    horizontal line at $z=1$~Mm; (top right) the total energy flux as
    a function of height;
    (center) the time-averaged magnetic field strength at $z=1$~Mm (black)
    and $z=0$~Mm (gray); and (bottom) the power spectrum of the
    field-aligned velocity, also showing the Gaussian 3-minute (5~mHz)
    and 5-minute (3~mHz) bands.}
  \label{auzpow}
\end{figure*}

The velocity field of this case is shown in Figures~\ref{vertBuz} and
\ref{vertBud}. As previously, the grayscale image shows the velocity,
the black contours outline the field concentrations at $z=0$~Mm,
now using a threshold of 300~G, while the white dashed line shows
the location of the periodic boundary. Figure~\ref{vertBuz} shows
the vertical velocity at $z=0.59$~Mm. At photospheric heights,
the velocity field is dominated by ``global'' oscillations with
large horizontal wavelengths and periods around 5 minutes.
At $z=0.59$~Mm, there are 
still some remnants of these, for example at
$x=13-17$~Mm after 2500~s, but the shocks generated in and around
the flux concentrations are stronger and have begun to fan out.
This figure is available as an animation in the online journal,
showing how the velocity field develops from $z=0$~Mm to $z=2$~Mm.
The animation shows quite clearly how the waves propagate upwards
and fan outwards.

Figure~\ref{vertBud} shows the velocity {\it along the magnetic field}
at $z=1$~Mm, and it is this velocity component we will be performing
our analysis on in this and all subsequent cases.\footnote{The sign of
the field-aligned velocity denotes whether it is parallel or anti-parallel
to the magnetic field. Since all our simulations have unipolar magnetic
fields, we take the direction generally corresponding to upward propagation
to be positive.} Since waves are forced to propagate
along the magnetic field in regions where the field is dominant, this
should give us a cleaner velocity signal than the vertical velocity;
the reason the field-aligned velocity was not used in case A is that
it is not a very meaningful quantity in weak-field (high-$\beta$) regions.

Unlike in case A, the velocity field at $z=1$~Mm
is now far from uniformly distributed.
As we see, all strong flux concentrations are associated with a region of high
velocity amplitude higher up, and nearly all regions of high amplitude
are connected with flux concentrations. One main exception is the region
of relatively high-frequency signal that starts out at $x=4$~Mm,
eventually moving to $x=6$~Mm. This area is situated between two
flux concentrations and is a region of interference between
waves coming from either side --- one can see the interference pattern
developing in the animated version of Figure~\ref{vertBuz}, which shows
the evolution of the wave pattern with height.
Some other regions between flux
concentrations, for example around $x=11.5$~Mm and for the last 1000~s
around $x=4$~Mm, show quite weak signals. The many curved wavefronts are
a consequence of the outward propagation of waves from the flux concentrations;
particularly notable examples are found at $x=2$~Mm, $x=6$~Mm, and
near the boundary at $x=16.5$~Mm. It is also interesting to note how much
the flux concentrations are moved around by the convective motion --- one
flux tube starts out at $x=2$~Mm and ends up at $x=6$~Mm, and several
flux concentrations merge during the simulation. This horizontal movement
is on the order of 1~km~s$^{-1}$, which is within the ranges reported
from observations of the movement of bright points in intergranular lanes
\citep{Berger+Title1996}.

\begin{figure*}
\begin{center}
  \includegraphics[scale=0.6]{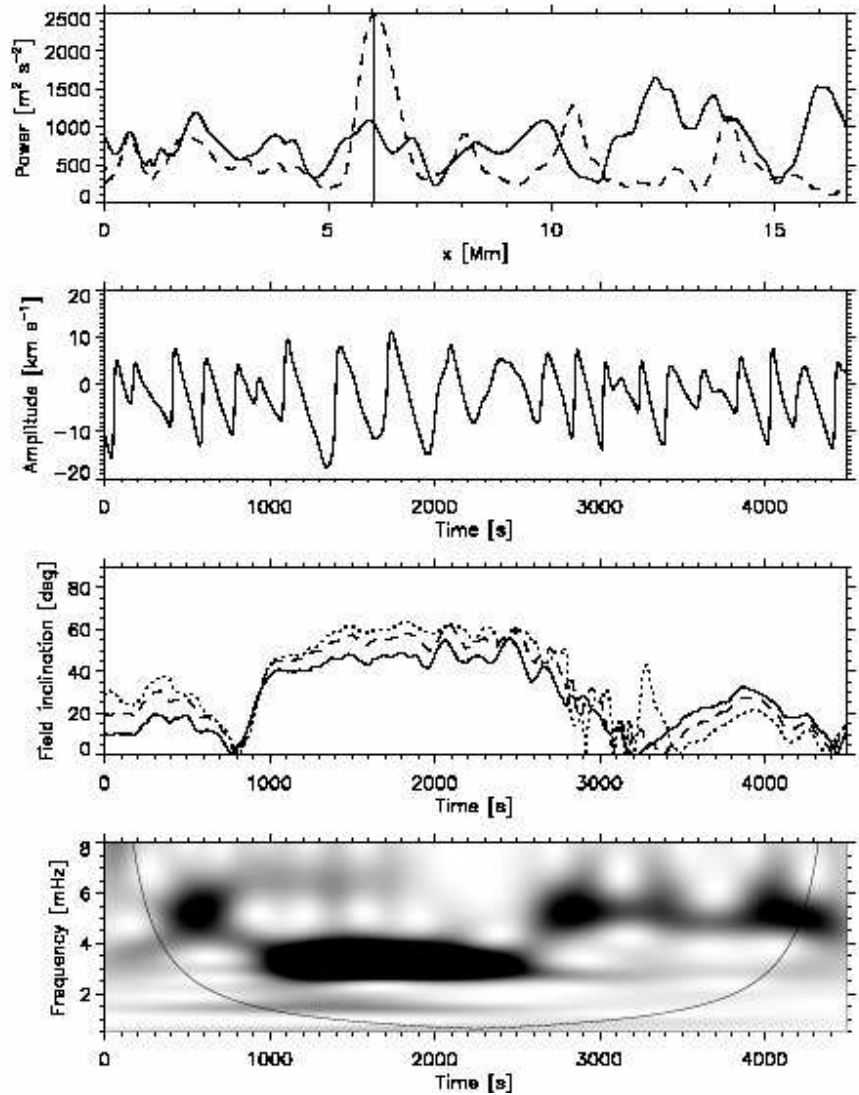}
  \caption{Power (top panel) in the 3-minute (solid line) and 5-minute
    (dashed line) bands at $z=1$~Mm in case B,
    with a vertical line marking the horizontal location where
    the other data are plotted; the field-aligned velocity signal
    (upper middle panel); the magnetic field inclination (lower middle
    panel; at $z=1$~Mm, solid; 250~km below along the field, dashed;
    500~km below along the field,
    dotted); and the wavelet power spectrum of the velocity (bottom panel).
    The horizontal position is $x=6.01$~Mm.}
  \label{vertpwrudang185}
\end{center}
\end{figure*}

\begin{figure*}
\begin{center}
  \includegraphics[scale=0.6]{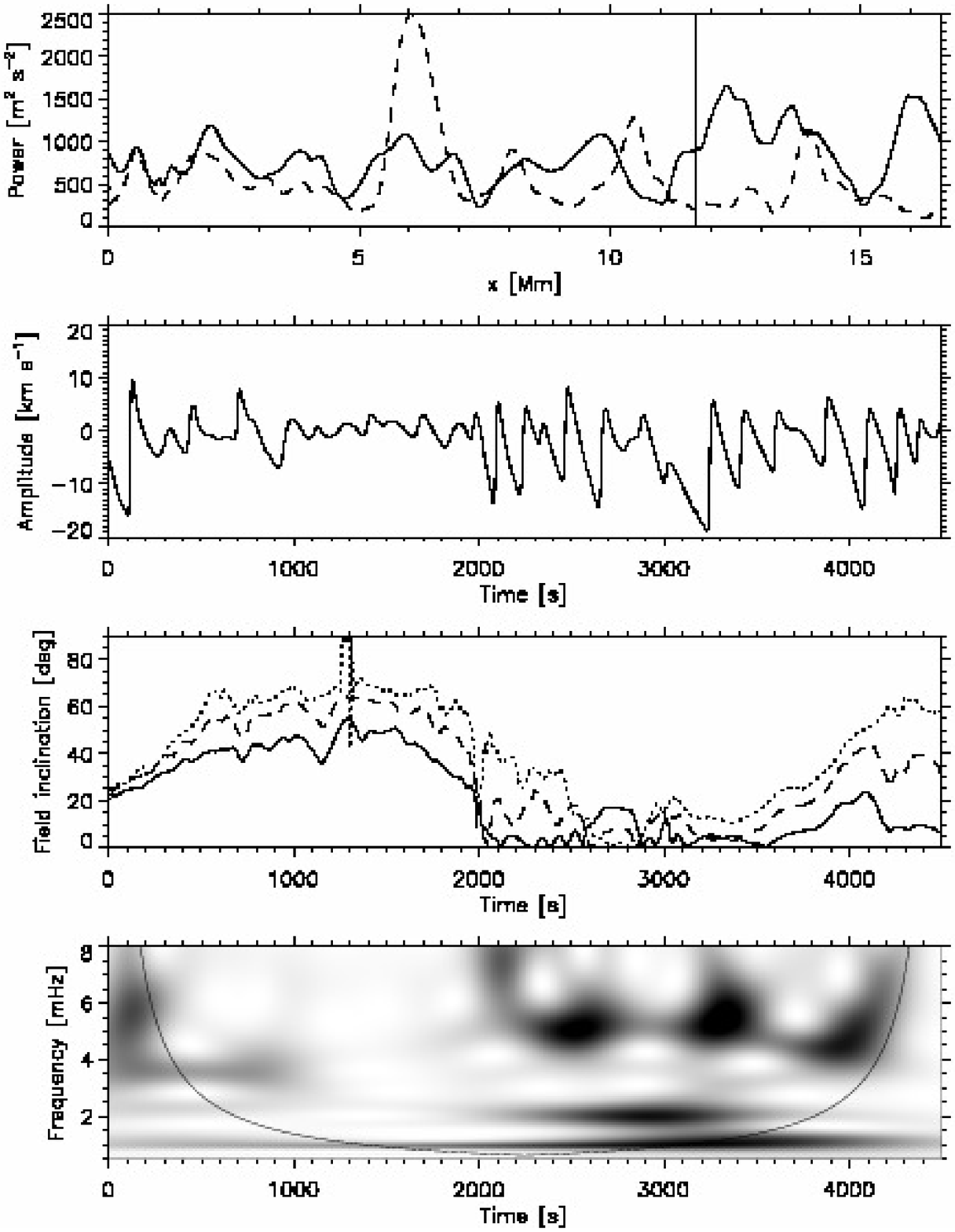}
  \caption{As Figure~\ref{vertpwrudang185}, but at $x=11.70$~Mm.}
  \label{vertpwrudang360}
\end{center}
\end{figure*}

As in case A, we have performed a Fourier analysis to investigate
the periodicity of the velocity signal. The results are shown in
Figure~\ref{auzpow}, which, like Figure~\ref{cuzpow} for case A,
shows the power spectrum in the bottom panel, the time-averaged
magnetic field in the middle panel, and the ratio of power in
the 3-minute and 5-minute bands in the top panel. As before,
the horizontal line in the top panels shows the analysis height
($z=1$~Mm), and the dark gray lines are time-averaged field
lines; the new thick white line indicates the time-averaged
height where $\beta=1$, the dividing line between the regions
dominated by the magnetic field (above the line) and by gas
pressure forces (below it). Although the central flux tube appears
very dominant when looking at the time-averaged field strength,
it is in fact not that much stronger than the others; however,
it moves much less horizontally, making it stronger when averaging
over the whole simulation.

The top panel shows that the 3-minute band remains dominant in the
chromosphere. This is as expected from classical theory, since the
acoustic cutoff period
is shortest there, and 5-minute disturbances are expected to be evanescent.
It is notable that the peaks in the
3-minute power seem to correspond very well with the peaks in the
magnetic field strength. If the effects of the reduced radiative relaxation
time in strongly magnetized regions were important, these are
locations where 5-minute propagation would be expected.
 
There are,
however, several windows in the chromosphere where there is also
significant power in the 5-minute band, notably around $x=6$~Mm,
$x=8$~Mm, and $x=10.5$~Mm.
The latter two are regions of 
inclined field on either side of the central strong flux
concentration. This fits
well with the results of many previous simulations \citep{Suematsu1990,
DePontieu+etal2004,DePontieu+etal2005,Heggland+etal2007}
that have found that otherwise evanescent long-period disturbances can
still propagate upwards along strong, inclined field, since the reduced
effective gravity increases the acoustic cutoff period. We should point
out that the previous simulations have generally assumed strong,
inclined flux tubes all the way down to the photosphere, while we find
that long-period propagation occurs even when only a part of the
propagation is along strong, inclined field at the edges of
a mostly vertical flux concentration.

On the other
hand, $x=6$~Mm is a region where the average field inclination is not
particularly high, at least not compared to the regions surrounding it.
However, as we have seen from Figure~\ref{vertBud}, the field
is quite dynamic, and the time-averaged field inclination shown in
the top panel of Figure~\ref{auzpow} does not tell
the whole story. In fact, a look at Figure~\ref{vertBud} shows that
$x=6$~Mm connects to two different photospheric flux concentrations
over the course of the simulation, and it seems to be particularly
during the time when the region is on the outer edge of the central flux tube
(from 1000 to 2500~s) that it is dominated by longer-period waves.
So we need to take the variations with time into account, not just
the averages over the whole simulation.

The Fourier analysis deals with averages over the whole analyzed time
period, and in order to get sufficient spectral resolution, one typically
needs time series on the order of one hour. The solar atmosphere, however,
is dynamic on timescales much shorter than that, and when conditions
at one location change significantly during the analysis time, the
Fourier analysis averages this out and can be
misleading. A wavelet analysis, which takes variations in time into
account, is a more precise tool; we have performed such an analysis
at some selected locations.

In Figure~\ref{vertpwrudang185}, the top panel shows the power in the 3
and 5-minute bands taken from the Fourier analysis,
with a vertical line marking the horizontal location, $x=6$~Mm, where
we are plotting the data in the three lower panels. The upper middle panel
shows the time series of the field-aligned velocity, while the lower middle
panel shows the variation of the (absolute) field inclination\footnote{We use the convention that an inclination of 0$^{\circ}$ means the field is vertical.}, both at
$z=1$~Mm. Since the appearance of the wave pattern does not just
depend on conditions at this height, but is a function of the conditions
throughout the lower atmosphere, we have also traced the field lines
downwards at each timestep and plotted, in the same panel, the inclination
250~km (dashed line) and 500~km (dotted line) lower along the field.
The bottom panel shows the wavelet power spectrum of the velocity
signal at $z=1$~Mm, calculated using the Morlet wavelet
\citep[see][]{Torrence+Compo1998}, with a black line
showing the cone of influence; values below this line are
subject to edge effects. Darker color corresponds to higher power.

In these plots we see how the period of the
velocity signal increases significantly between about 1000 and 2600~s,
which is the time when this region is located towards the outer edge
of the central flux concentration
and the inclination is much greater than earlier and later. At
this time we get a large peak in the power spectrum at periods around 5 minutes,
while at other times, most of the power is in 3-minute waves.
This again strongly supports the results in the earlier literature,
showing that field inclination is important for long-period wave
propagation.

So, do inclination increases always result in increased 5-minute power?
Yes and no. The peaks in the 5-minute power tend to be associated with
periods of increased inclination, but the area around $x=6$~Mm is
unique in having such a strong, relatively pure signal coherent over
five wave periods; hence the large peak in the power spectrum there. But
even in other locations where the 5-minute power does not increase
markedly, the wave dynamics are still quite different when the
inclination is large. An example can be seen in Figure~\ref{vertpwrudang360},
showing the velocity signal and inclination at $x=11.70$~Mm. Here,
the inclination is large for the first 2000~s, and the velocity signal
at that time is weaker and more irregular than later in the simulation,
when the field is more vertical. One explanation for this can
be deduced from a comparison with Figure~\ref{vertBud}: the times
when the inclination is large are, in general, the times when the
region is located at the interface between two flux concentrations.
Since the flux concentrations are the main sources of wave power
in the system, such interface regions are also regions of interference
between the waves coming from the two flux concentrations. In many
cases, this interference is destructive and reduces the amplitude of
the disturbances. The periodicity is also perturbed since the signal is now
a superposition of the waves coming from the two sources. Examples of
regions with such destructive interference are at $x=4$~Mm after 3000~s
(a case of triple interference, as there is a weaker, $B_z < 300$~G 
flux concentration acting as a wave source below this location),
and the plotted $x=11.7$~Mm for the first 2000~s. In contrast, the
regions of most dominant 5-minute power, at $x=6$~Mm and $x=10.5$~Mm, are
located towards the sides of the region dominated by the strongest
central flux tube, with the interference regions located on the outside
of these.

It is also worth noting that the regions of strongest 3-minute dominance in
the chromosphere are at $x=12-13$~Mm and $x=16$~Mm, both of which are
directly above flux concentrations that show relatively little horizontal
movement (cf. Figure~\ref{vertBud}).

\subsection{Case C}

\begin{figure}
\begin{center}
  \includegraphics[scale=0.45]{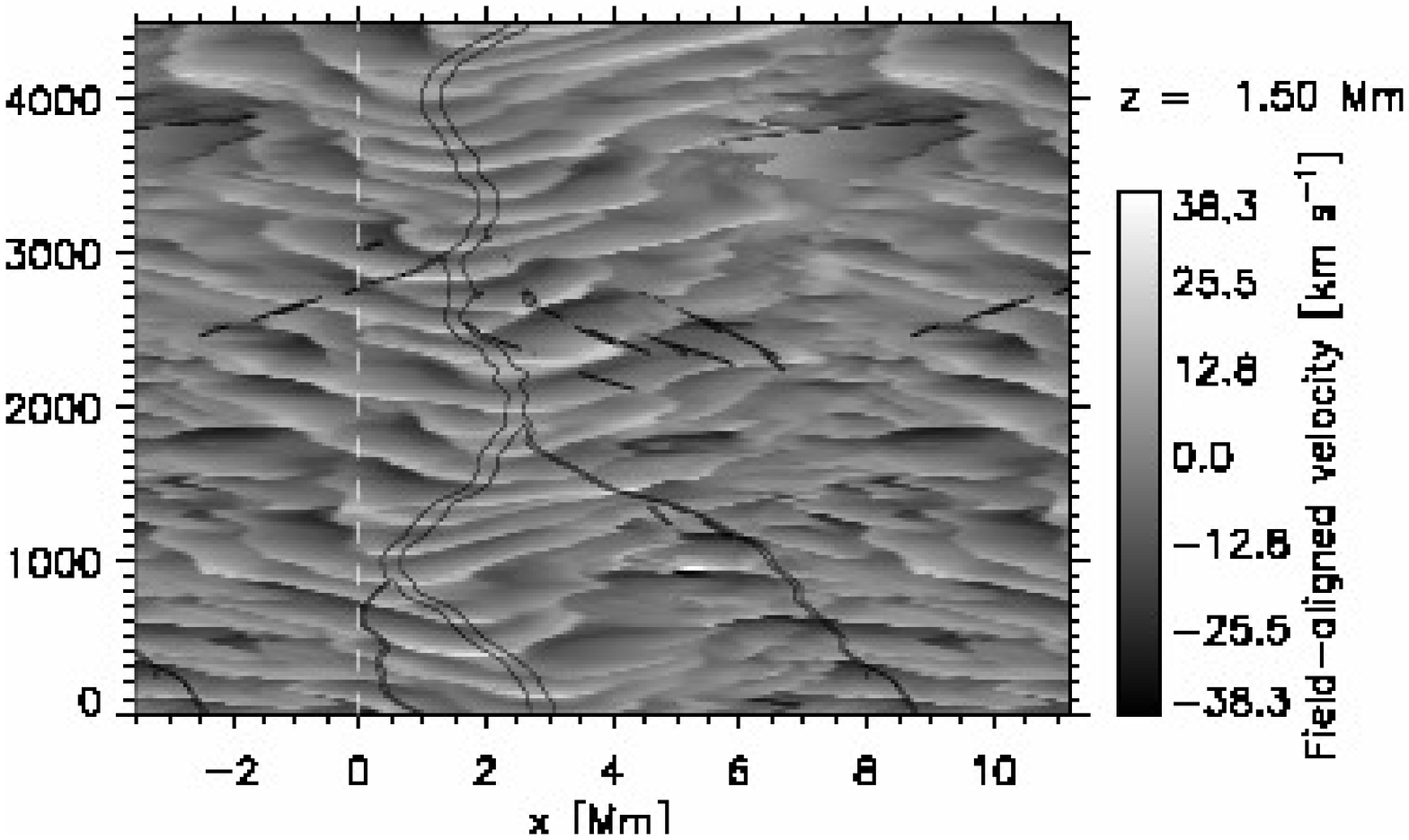} \\
  \includegraphics[scale=0.45]{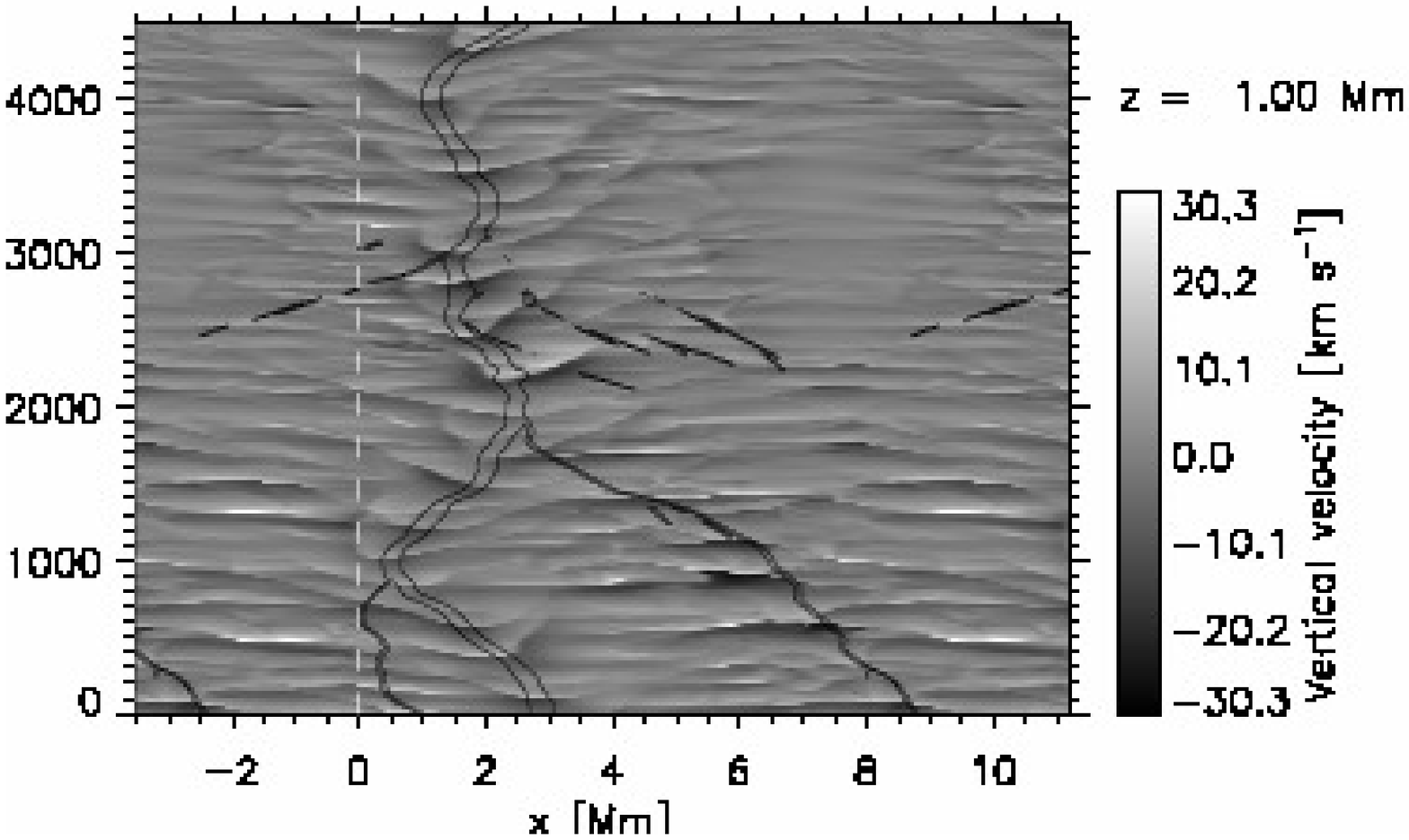}
  \caption{Field-aligned velocity at $z=1.5$~Mm (top panel) and
    vertical velocity at $z=1$~Mm (bottom panel) in case C, with
    superimposed contours showing the regions where $B_z$ at
    $z=0$~Mm is greater than 150~G. The white dashed line shows
    the location of the periodic boundary; the rightmost $3.5$~Mm
    are repeated on the left.}
  \label{poreABuz}
\end{center}
\end{figure}

\begin{figure*}
  \plotone{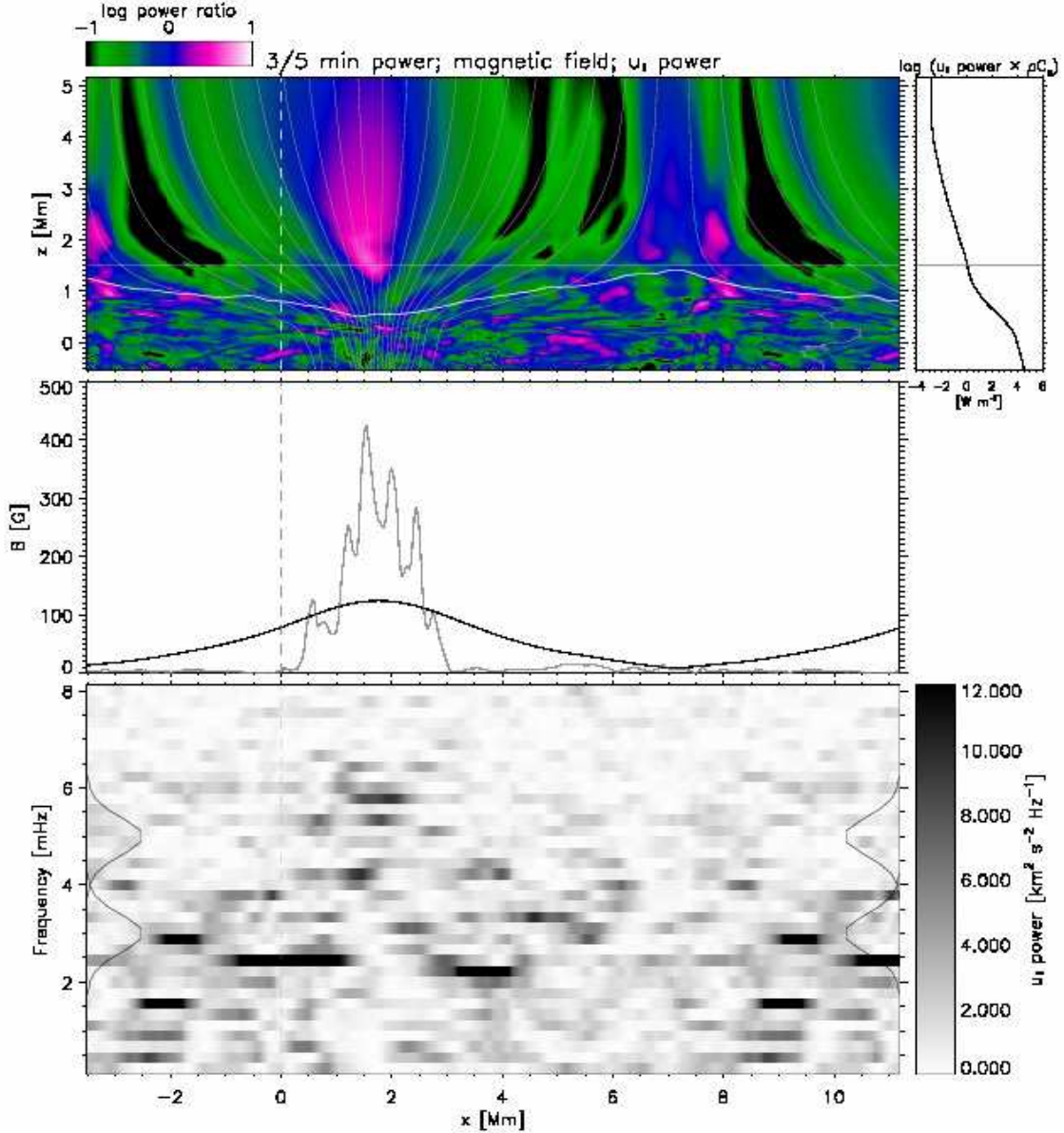}
  \caption{Fourier analysis of case C, showing (top) the ratio of 3-minute
    power to 5-minute power, with time-averaged $\beta$=1 surface (thick
    white line) and
    magnetic field lines (thin gray lines) superimposed, as well as a
    horizontal line at $z=1.5$~Mm; (top right) the total energy flux as
    a function of height;
    (center) the time-averaged magnetic field strength at $z=1.5$~Mm (black)
    and $z=0$~Mm (gray); and (bottom) the power spectrum of the
    field-aligned velocity, also showing the Gaussian 3-minute (5~mHz)
    and 5-minute (3~mHz) bands. The vertical dashed line shows the location
    of the horizontal boundary at $x=0$~Mm.}
  \label{poreAfourier}
\end{figure*}

While case B had several magnetic flux concentrations of comparable strength,
case C contains only one dominant flux tube. Unhindered by the influence of
neighboring flux tubes, it can expand freely and fill the entire simulation
box at coronal heights. This expansion creates a large region of inclined
field at the edges of the flux concentration, which was where we preferentially
saw 5-minute propagation in case B. The initial state of this case is shown
in the middle panel of Figure~\ref{initfig}.

Although the main flux concentration is strong, with a $\beta=1$ height
reaching down to $z=-0.5$~Mm (far below the photosphere) at any given time,
the magnetic field in the
rest of this model is rather weak. As a result, the average height where
$\beta=1$ is higher in this case than in case B (see Figure~\ref{poreAfourier});
between $x=5$~Mm and $x=8$~Mm, it is above our analysis height of 1~Mm,
and the field-aligned velocity is not a meaningful quantity there. Because of
this, we will be performing the analysis of this case at $z=1.5$~Mm.
Figure~\ref{poreABuz} (top) shows the field-aligned velocity at that height 
(grayscale),
with superimposed black contours showing where $B_z$ at $z=0$~Mm is greater
than 150~G. This value was chosen to point out the secondary flux
concentration that merges with the primary at $t=1900$~s; the primary has
a field strength on the order of 2000~G. Since the primary flux concentration
is located towards the left of the simulation box, we have plotted the rightmost
3.5~Mm again on the left, with the white dashed line marking the location of
the boundary.

Although there are some powerful shocks connected with the secondary flux
concentration in the first 1200~s, the velocity field is dominated by the
waves propagating outwards from the primary. New oscillations are constantly
being generated in the center of the flux concentration.
(Figure~\ref{propagation} is an example of such wave propagation,
taken from this case; see also \citeauthor{Kato+etal2011}
\citeyear{Kato+etal2011} for a study of wave generation in a similar
model.) These propagate upwards
and outwards, but there is large variability in how fast they move, and how far 
to the sides they reach. In general, the waves travel faster and longer in the 
direction that the flux concentration itself is moving at the time; this is not
unreasonable,
as the waves from the lower layers will tend to align with the axis of the flux
concentration, and this is itself inclined in the direction the flux
concentration
is moving. We do occasionally see some wavefront merging as a result of these
varying propagation patterns, and this could potentially affect the calculated
periodicities; however, most of the wavefronts ``absorbed'' this way are rather
weak. Two things are obvious from the picture: although the flux concentration
is only a few hundred km wide at photospheric heights, it dominates a very large
area (up to 4~Mm to each side) in the upper chromosphere, and the period of
oscillations at its center is significantly shorter than that of the
oscillations to the side.

Since the analysis in the other cases is performed at $z=1$~Mm, we also
show the velocity field at that height in this case (Figure~\ref{poreABuz},
bottom panel). As $z=1$~Mm is partially below
the average $\beta=1$ height, we here plot the vertical velocity rather
than the velocity along the magnetic field.
In this case, we see wavefronts that are coherent over several Mm
horizontal distance, especially in the first 2500~s. These sometimes
become very strong shocks with amplitudes of more than 30~km~s$^{-1}$, 
and some of the strongest are connected with the secondary flux
concentration. The typical period of these oscillations is 200-250~s,
though there is some variation. Meanwhile, in part superimposed
on the larger ``global'' pattern, we see the wavefronts clearly propagating
outwards from the primary flux concentration. The area farthest from the
flux concentration, around $x=8$~Mm, shows very weak signal in the
later part of the simulation. At this time, the magnetic field in the region is
close to horizontal, and temporarily lower temperature makes the medium
magnetically dominated. This creates a ``lid'' which inhibits vertical wave
propagation.

As in the other cases, we have performed a Fourier analysis, the results
of which are shown in Figure~\ref{poreAfourier}. 
A pattern is clear: we see 3-minute propagation
mainly above the center of the flux concentration, and to some extent in the
very weak-field area around $x=6-8$~Mm, and 5-minute propagation in the
inclined field regions on the sides of the flux concentration. Although
the ratio plot shows this even more clearly at greater heights, it is
also plainly visible in the
spectrum itself; the central flux tube region between $x=1.5$~Mm and $x=2.5$~Mm
has its whole spectrum shifted towards higher frequencies than in the regions
surrounding it. This is in very good agreement with what we found in case B.

\subsection{Case D}

\begin{figure}
\includegraphics[scale=0.43]{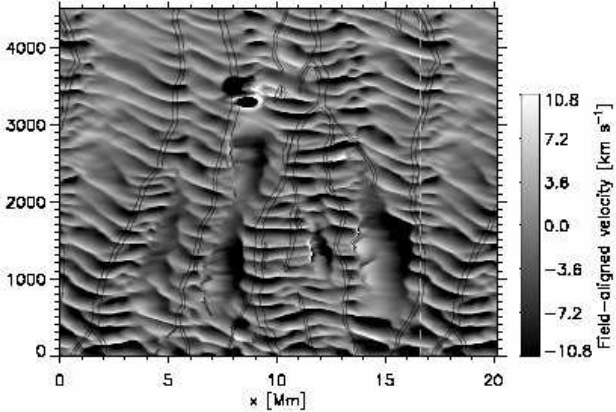}
  \caption{Field-aligned velocity at $z=1$~Mm in case D, with
    superimposed contours showing the regions where $B_z$ at
    $z=0$~Mm is greater than 300~G. The white dashed line shows
    the location of the periodic boundary; the leftmost $3.5$~Mm
    are repeated on the right.}
  \label{45degBud}
\end{figure}

\begin{figure*}[tbp]
  \plotone{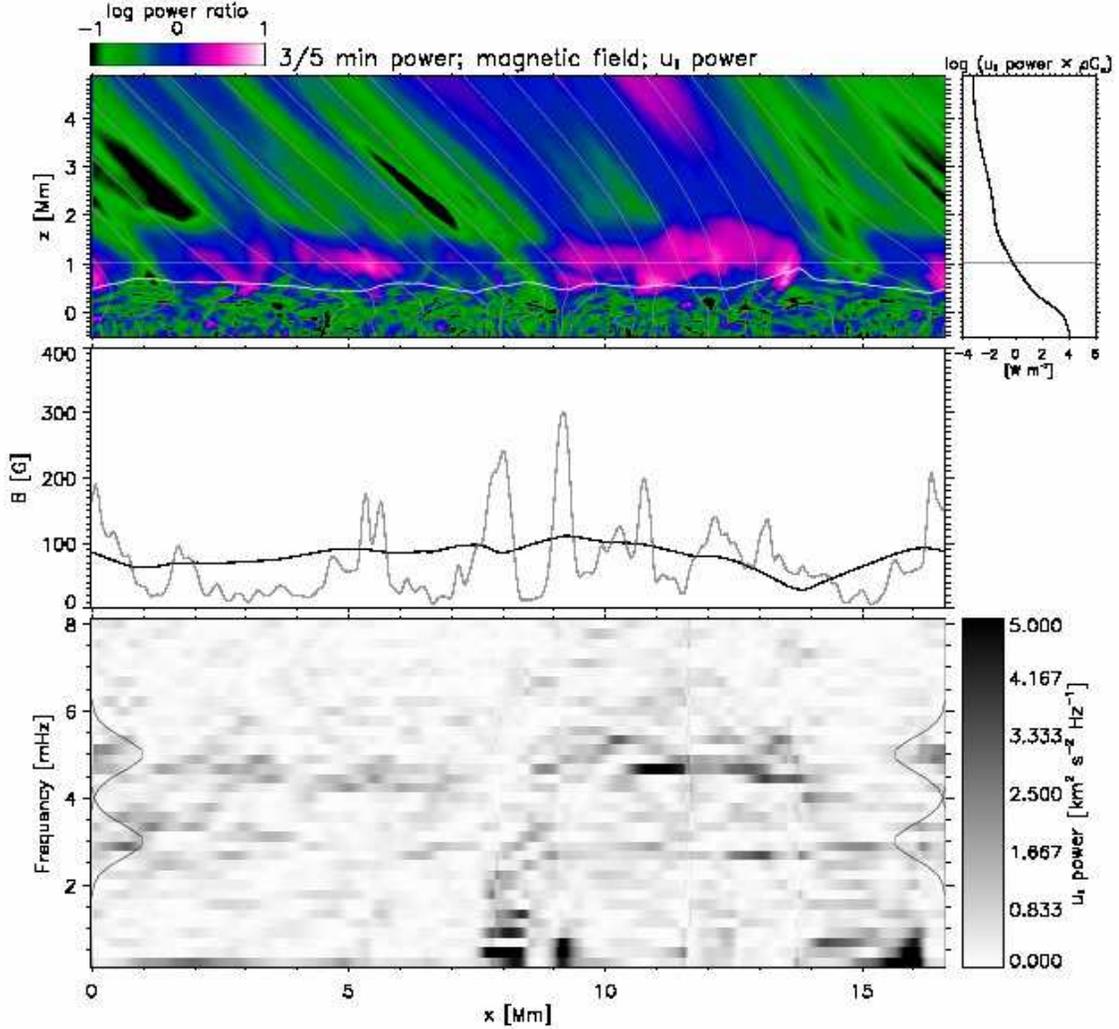}
  \caption{Fourier analysis of case D, showing (top) the ratio of 3-minute
    power to 5-minute power, with time-averaged $\beta$=1 surface (thick
    white line) and
    magnetic field lines (thin gray lines) superimposed, as well as a
    horizontal line at $z=1$~Mm; (top right) the total energy flux as
    a function of height;
    (center) the time-averaged magnetic field strength at $z=1$~Mm (black)
    and $z=0$~Mm (gray); and (bottom) the power spectrum of the
    field-aligned velocity, also showing the Gaussian 3-minute (5~mHz)
    and 5-minute (3~mHz) bands.}
  \label{budpow}
\end{figure*}

We now turn our attention to case D, which has similar flux concentrations
as case B at the photospheric level, but also includes an additional imposed
uniform field of $45^{\circ}$ inclination (slanting towards lower $x$).
This uniform field is dominant in the corona and upper chromosphere,
but is not generally strong enough to dominate dynamics at the
photospheric or lower chromospheric levels.
The initial state of this model is shown in the second panel from the bottom
of Figure~\ref{initfig}.
This case is meant to represent a plage region with nearby field of
opposite polarity.

Figure~\ref{45degBud}, like Figure~\ref{vertBud} for case B, shows the
field-aligned velocity at $z=1$~Mm in case D as a function of $x$ and time,
with overplotted black flux concentrations ($B_z > 300$~G) at $z=0$~Mm
and the leftmost
$3.5$~Mm plotted again on the right due to the flux concentration on the
boundary. As in case B, we see that the regions with a powerful velocity
signal are correlated with the regions of strong flux concentrations
below. In fact, the association seems even stronger in case D, with large
areas between flux concentrations showing quite weak velocity signals
(e.g. between $x=2$ and $x=4$ Mm in the later half of the simulation time).
Other regions between flux concentrations, e.g. at $x=8$~Mm and $x=15$~Mm,
become sites of less regular flows, especially early in the simulation. The
field inclination in these regions at those times is generally very large,
even near-horizontal in the region around $x=15$~Mm. The sudden outburst
at $x=8.5$~Mm between 3200~s and 3500~s is due to a reconnection event
at $z \approx 0.5$~Mm. In general, due to the left-leaning field, each
flux concentration dominates the velocity in a large area to its left,
but only a short distance to its right.

A Fourier analysis of this case is shown in Figure~\ref{budpow}.
As in case B, the 3-minute band is dominant in large parts of the
chromosphere (notably $x=2-6$~Mm and $x=9-14$~Mm), but there are some
windows where the 5-minute band
dominates. The most prominent of these are around $x=1$~Mm, $x=8$~Mm,
and $x=15$~Mm. The latter two also exhibit large amounts of low-frequency
power, due to the non-periodic flows present there.

Compared to case B (Figure~\ref{auzpow}), there is less of a
connection between the strongest average photospheric fields and the peaks in
the 3-minute power, even allowing for the slight leftward drift
that would be expected due to the inclination of the additional
homogeneous field. This is likely primarily because the
flux concentrations in this model display more horizontal movement
than in case B (cf. Figure~\ref{budpow}) --- for example, the strong
flux tube that starts out at $x=10$~Mm moves first left to $x=9$~Mm,
and later right to $x=12$~Mm. Meanwhile, there is a separate flux
concentration at $x=13-14$~Mm that eventually merges with the other
one after 3000~s. The area connected to these is in general the
area of greatest velocity power, in particular in the 3-minute band.
It is also the region of lowest average field inclination.
The flux concentration at the boundary, $x=16.5$~Mm, moves comparatively
little and this region does correspond to a peak in the 3-minute power.
Overall, though, comparison with the average fields and inclinations
is of less value in case D due to the horizontal movement. In order
to figure out what is going on, it is necessary to take time variations
into account. The wavelet
analysis is a tool better suited for this analysis, and we will
again look at the time variation of the velocity and
inclination at some specific horizontal positions.

\begin{figure*}
\begin{center}
  \includegraphics[scale=0.6]{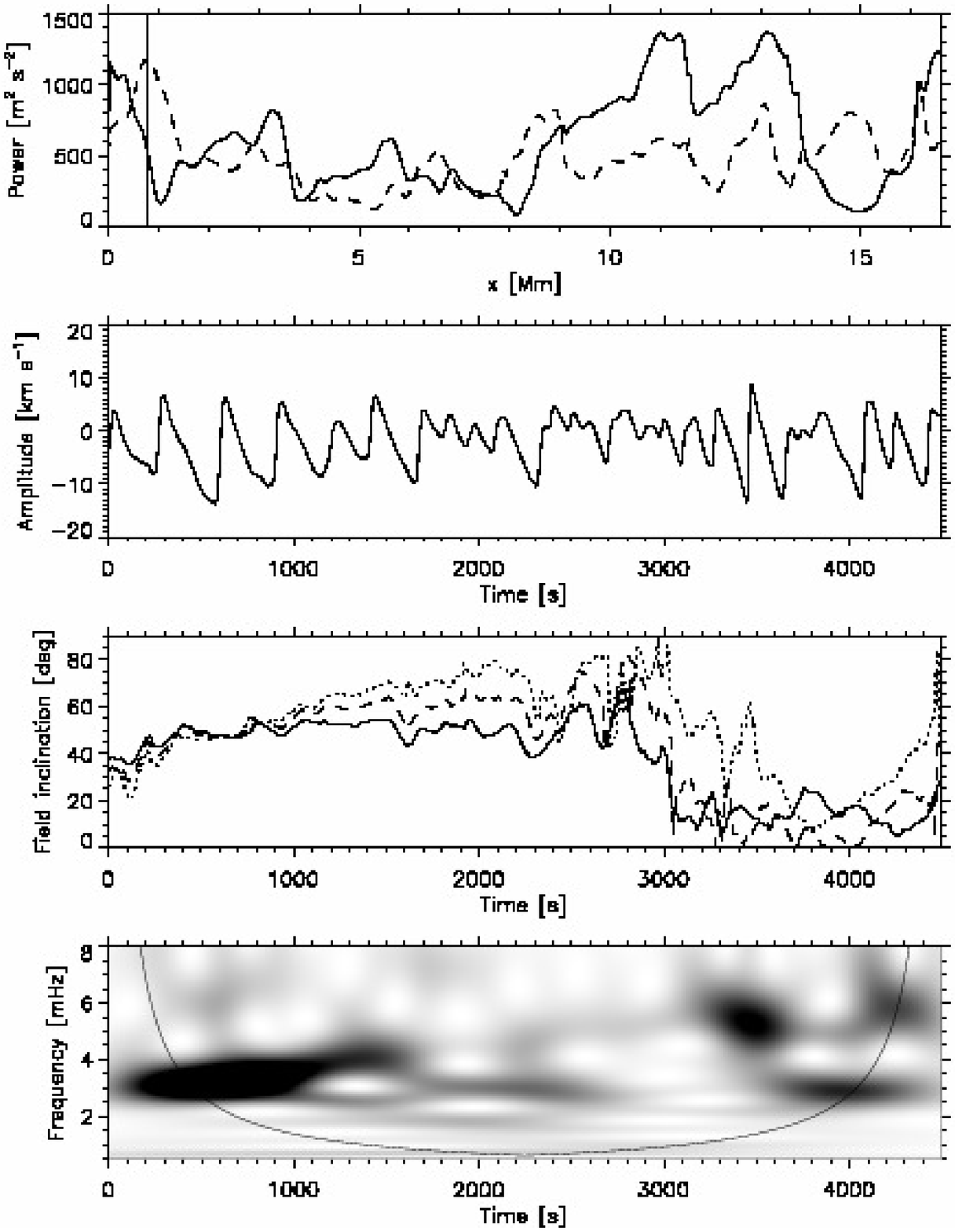}
  \caption{Power (top panel) in the 3-minute (solid line) and 5-minute
    (dashed line) bands at $z=1$~Mm in case D,
    with a vertical line marking the horizontal location where
    the other data are plotted; the field-aligned velocity signal
    (upper middle panel); the magnetic field inclination (lower middle
    panel; at $z=1$~Mm, solid; 250 km lower along the field, dashed;
    500 km below along the field,
    dotted); and the wavelet power spectrum of the velocity (bottom panel).
    The horizontal position is $x=0.78$~Mm.}
  \label{bampang024}
\end{center}
\end{figure*}

\begin{figure*}
\begin{center}
  \includegraphics[scale=0.6]{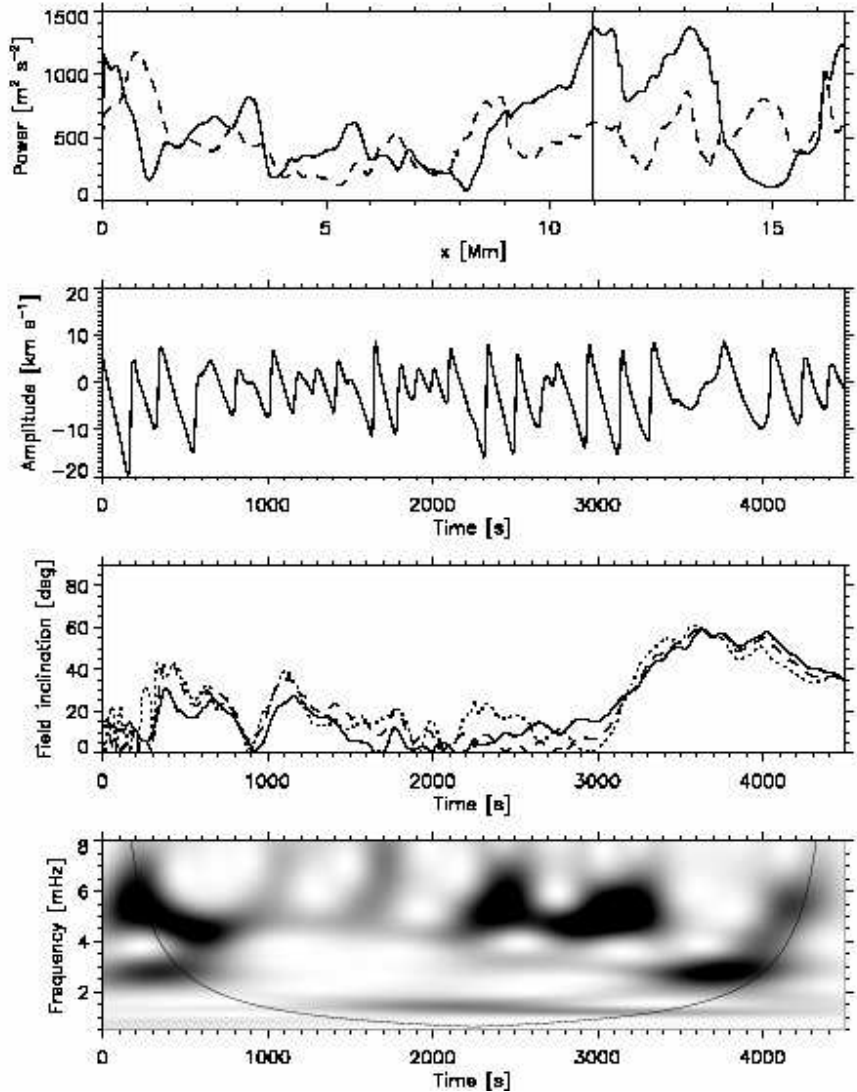}
  \caption{As Figure~\ref{bampang024}, but at $x=10.99$~Mm.}
  \label{bampang338}
\end{center}
\end{figure*}

\begin{figure*}
\begin{center}
  \includegraphics[scale=0.6]{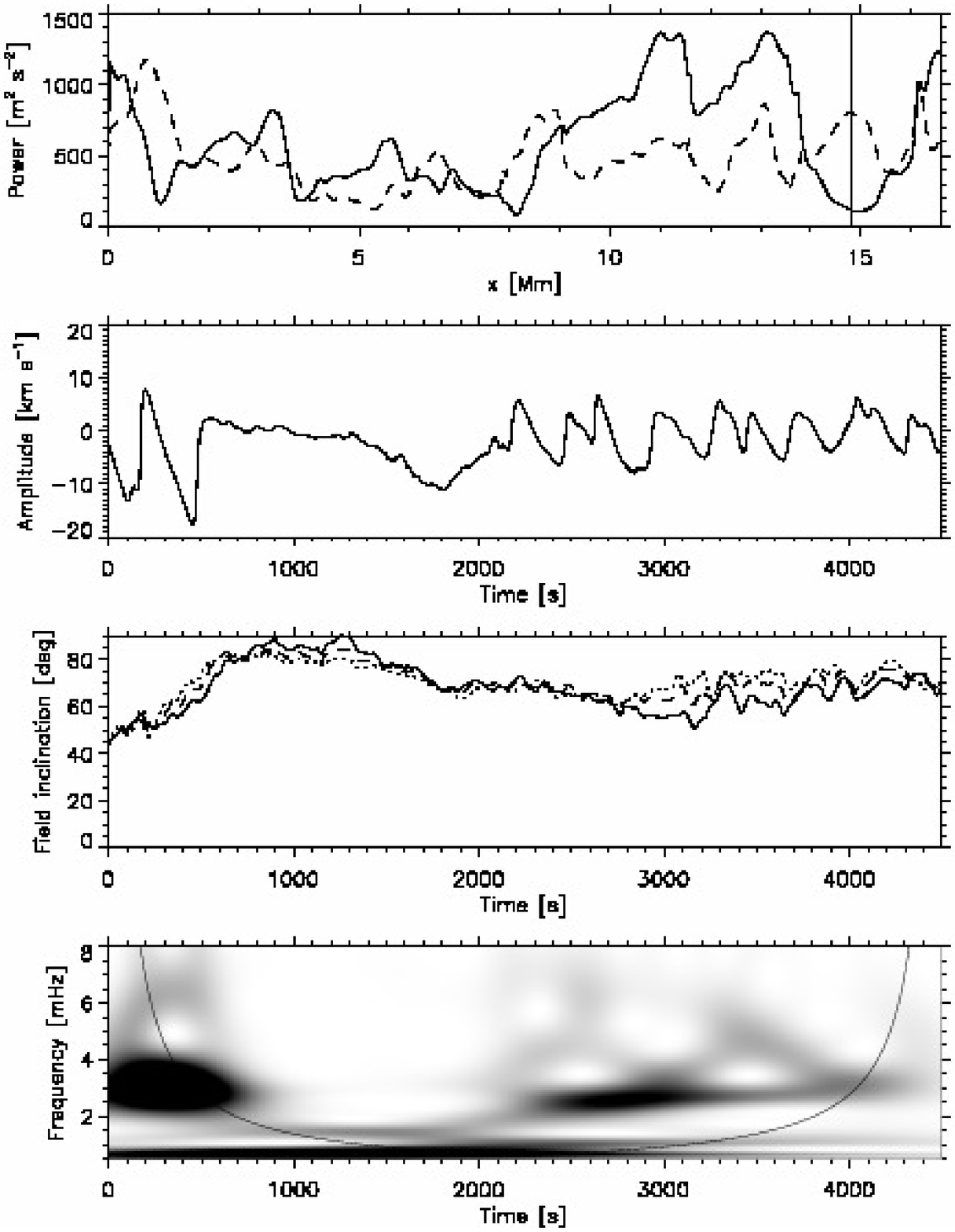}
  \caption{As Figure~\ref{bampang024}, but at $x=14.82$~Mm.}
  \label{bampang456}
\end{center}
\end{figure*}

Figure~\ref{bampang024} is a four-panel figure showing (top) the
3-minute (solid) and 5-minute (dashed) power, with a vertical line
showing the current horizontal location ($x=0.78$~Mm); (upper middle)
the field-aligned velocity as a function of time at that location;
(lower middle) the field inclination as a function of time at that location;
and (bottom) the wavelet power spectrum.
All panels are plotted at the height $z=1$~Mm; as previously, the field
inclination is also shown at two lower heights along the field.
The region around $x=0.78$~Mm has the
highest 5-minute power in the simulation, and it appears to be mainly
due to the behavior in the first 1700~s. At that time, there is a
quite regular train of long-period shock waves coming up. Between
1700~s and 3000~s, the velocity signal is much more irregular,
correlated with large and rapid variations in the inclination
at lower heights, although the dominant
periodicity remains 5~minutes. A comparison with Figure~\ref{45degBud}
shows that the change in behavior happens when the region becomes
an interference region between the flux concentration at the boundary
and the one starting out at $x=1.5$~Mm after roughly 1700~s; before
that, it is dominated by waves coming from the latter flux concentration.
After about 3300~s, the flux concentration at the boundary has moved to
a position just below our observation point; the inclination becomes
smaller and we get a wave train with shorter periodicity. The 
general behavior is similar to what we saw in case B; much of the
difference seems to be that with an imposed, left-leaning $45^{\circ}$
field, each flux concentration tends to dominate the wave field in
a large region to its left, but only in a small region to its right.

Figure~\ref{bampang338} shows the situation at $x=10.99$~Mm, one of the
largest peaks of 3-minute power. Here, the inclination remains mostly
below $20^{\circ}$ at $z=1$~Mm and $40^{\circ}$ at lower heights,
and we get a fairly regular short-period velocity
signal, though with varying amplitude. In the last 1000~s, the inclination
increases, and this does lead to a more irregular signal, with longer
time between the peaks, though the effect seems noticeable only during
the relatively short time when the inclination is above $40^{\circ}$.

Finally, Figure~\ref{bampang456} shows the situation at $x=14.82$~Mm,
at the center of a wide region where the 3-minute power is low and
the 5-minute power is dominant. This is a region where the field
inclination is generally very large --- during most of the simulation,
it is more than $60^{\circ}$. A few initial shocks, which actually
dominate the power spectrum, are followed by a long
period with near horizontal field and a steadily increasing downflow,
culminating at 1800~s. Then, after
2200~s, a rather irregular, mostly long-period signal sets in.
A comparison with Figure~\ref{budpow} shows that this signal comes
from the flux tube on the boundary to the right. A new flux concentration
appears (at $z=0$~Mm) after around 2000~s, but it connects only to the left
and does not dominate the velocity field directly above it. The most
notable effect at this location is the near complete absence of
3-minute signal due to the large field inclination.

In sum, this case shows a behavior that is rather similar to case B,
which has a more vertical field. We see 3-minute signal above the
strongest flux concentrations, where the vertical component of the
field is usually strong enough to keep the inclination low, and
5-minute signal towards the edges of the regions dominated by each
flux concentration. Among the differences between case B and case D,
we find that case D has less power at high frequencies, connected
with interference regions in case B, and that the 3-minute signal
is very weak in regions of heavily inclined field, which 
were absent in case B. The following case looks into what happens
when the strength of the constant inclined field is increased,
allowing it to dominate at lower heights.

\subsection{Case E}

\begin{figure}
\begin{center}
  \includegraphics[scale=0.43]{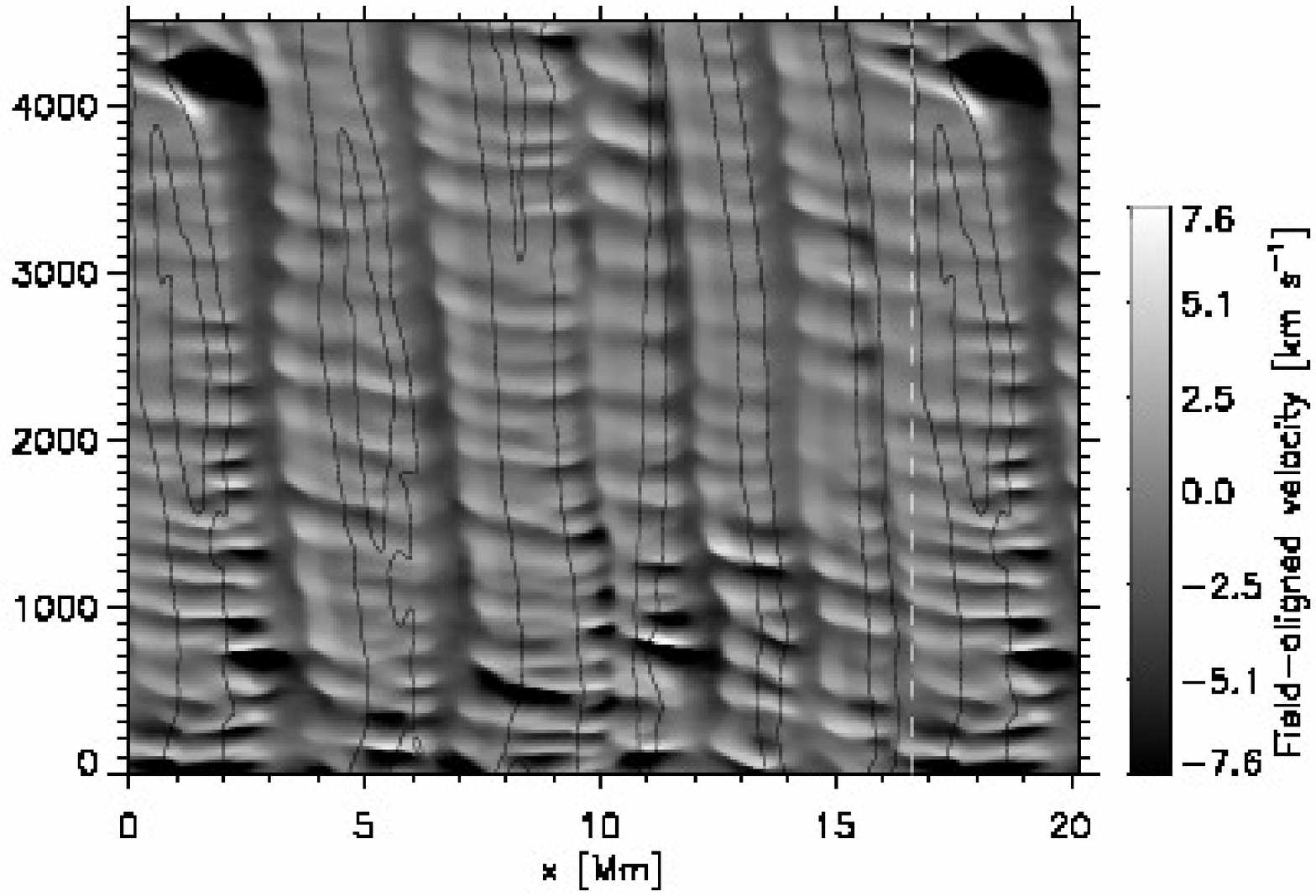}
  \caption{Field-aligned velocity at $z=1$~Mm in case E, with
    superimposed contours outlining the regions where $B_z$ at
    $z=0$~Mm is greater than 1000~G. The white dashed line shows
    the location of the periodic boundary; the leftmost $3.5$~Mm
    are repeated on the right.}
  \label{caseEBud}
\end{center}
\end{figure}

\begin{figure*}[tbp]
  \plotone{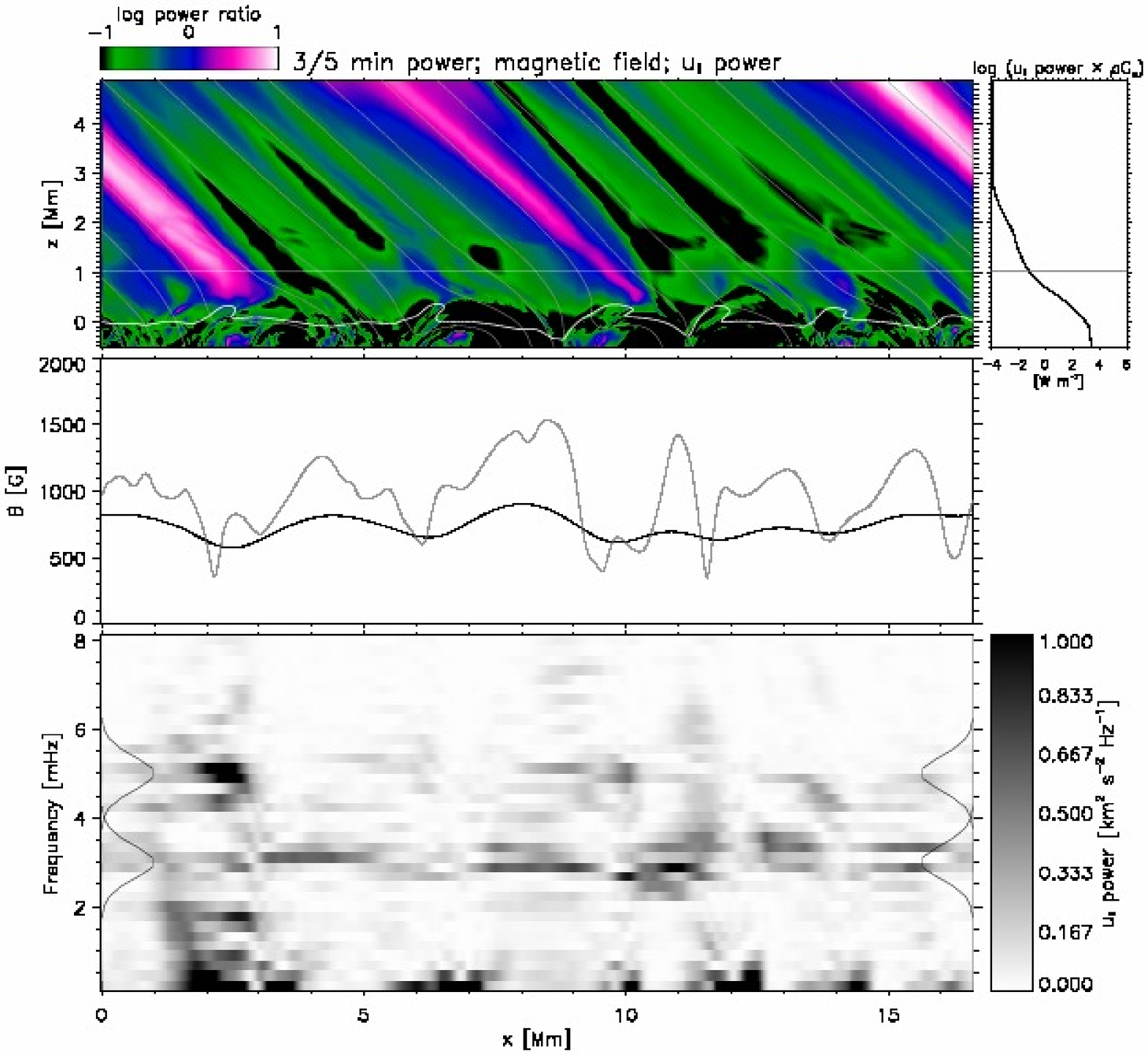}
  \caption{Fourier analysis of case E, showing (top) the ratio of 3-minute
    power to 5-minute power, with time-averaged $\beta$=1 surface (thick
    white line) and
    magnetic field lines (thin gray lines) superimposed, as well as a
    horizontal line at $z=1$~Mm; (top right) the total energy flux as
    a function of height;
    (center) the time-averaged magnetic field strength at $z=1$~Mm (black)
    and $z=0$~Mm (gray); and (bottom) the power spectrum of the
    field-aligned velocity, also showing the Gaussian 3-minute (5~mHz)
    and 5-minute (3~mHz) bands.}
  \label{caseEfourier}
\end{figure*}

In case E, the imposed field is still at a 45$^{\circ}$ angle with respect
to the vertical, but the field is significantly stronger than in case D.
In the mid-chromosphere, around $z=1$~Mm, it is about 8 times stronger;
lower down it is more variable.
The field-aligned velocity at $z=1$~Mm is plotted in
Figure~\ref{caseEBud}, with overplotted contours outlining the
regions where $B_z$ at $z=0$~Mm is greater than 1000~G (as compared to
the 300~G used in the similar plot for case D, Figure~\ref{45degBud}).
The initial state of the model is shown in the bottom panel of
Figure~\ref{initfig}. This model represents conditions in very strong
plage with nearby field of opposite polarity.

As in the other cases, the regions of highest velocity amplitude seem
to be located directly above the strongest flux concentrations. This
may seem counterintuitive at first --- with such a strong field acting
as a wave guide, we might expect the velocity signal to appear to the
side of the locations of the flux concentrations. But we should keep in
mind that the flux concentrations
themselves are fairly wide. Many have widths on the order of 1~Mm, or
the same as the height we are observing at, so even though the sideways
propagation starts around $z=0.5$~Mm (see Figure~\ref{caseEfourier}),
the waves coming from a flux concentration will still tend to be located
above it. As we move higher in the atmosphere,
the velocity signal does move sideways along the magnetic field, but
this effect is not yet very pronounced at $z=1$~Mm.

Between the flux concentrations, there are channels where almost no
wave signal can be seen. These neatly split the computational box into
zones of influence for each flux concentration.
The channels are actually downflow regions,
with slowly varying velocities of $1.5-2$~km~s$^{-1}$. A single larger
downflow event can be seen between $x=1$ and $x=3$~Mm after 4000~s.

Figure~\ref{caseEfourier} shows the Fourier analysis of this case. At
this field strength and inclination, the 5-minute band is dominant
across most of the chromosphere (top panel). Only
in two locations, around $x=2.4$~Mm and $x=10$~Mm, does the 3-minute band
dominate. The 3-minute power is highest
above high-$\beta$ regions, i.e. where the magnetic
field is weak. This is the opposite of what was found in case B (Figure
\ref{auzpow}), where the strongest flux concentrations were the sites
of highest 3-minute power. In that case, however, the strongest fields
were also the most vertical, whereas here, the weaker fields are more
vertical than the stronger fields. The results therefore support
the idea that the field inclination is central to wave propagation.

The full Fourier spectrum at $z=1$~Mm (bottom panel) shows the
same situation:
significant power in the 5-minute band across most locations, though
the 3-minute peak at $x=2.4$~Mm is the strongest in the simulation.
The 5-minute power is more evenly
distributed, with one broad peak at $x=11$~Mm and some smaller, also
quite broad peaks. We
also see the very low-frequency power connected with the downflow regions.

\begin{figure*}
\begin{center}
  \includegraphics[scale=0.6]{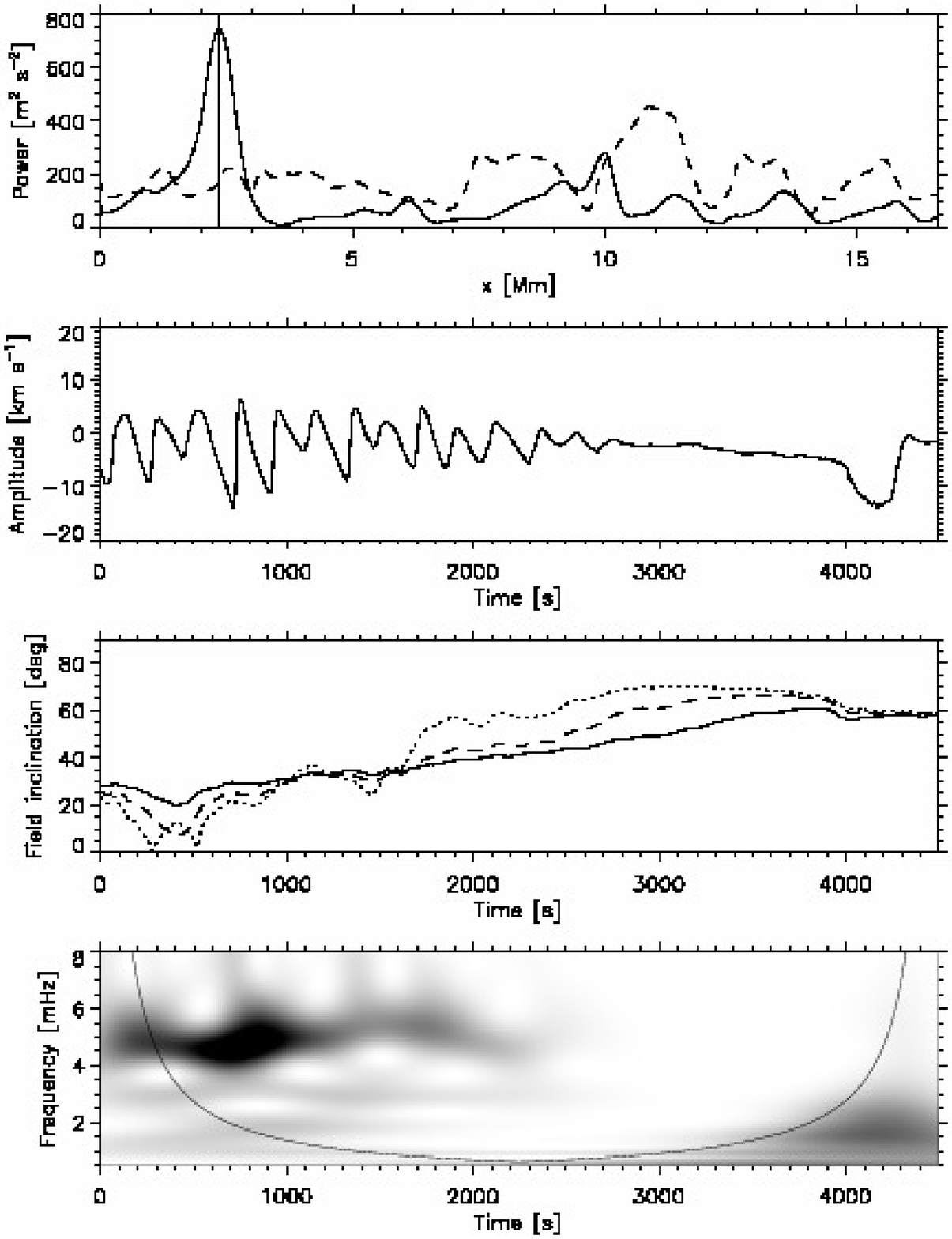}
  \caption{Power (top panel) in the 3-minute (solid line) and 5-minute
    (dashed line) bands at $z=1$~Mm in case E,
    with a vertical line marking the horizontal location where
    the other data are plotted; the field-aligned velocity signal
    (upper middle panel); the magnetic field inclination (lower middle
    panel; at $z=1$~Mm, solid; 250~km below along the field, dashed;
    500~km below along the field,
    dotted); and the wavelet power spectrum of the velocity (bottom panel).
    The horizontal position is $x=2.37$~Mm.}
  \label{Ewvl073}
\end{center}
\end{figure*}

Although the strong-field case E has a more steady velocity
field than the cases with weaker fields, there is still some drift
and variation with time. We once again perform a wavelet analysis
in order to look more closely at the wave propagation.
Figure~\ref{Ewvl073} shows the results at $x=2.37$~Mm, the largest
peak of 3-minute power. The velocity amplitude (upper middle panel)
is large in the initial stages, when the inclination is relatively
low (less than or around 30$^{\circ}$). There is power at a rather
wide range of frequencies at that time, though by far the strongest
signal is found in the 3-minute band between 4 and 6~mHz. After
2000~s, most of the signal dies out, as this vertical location
becomes part of a rather quiescent downflow region with velocities
of 4-5~km~s$^{-1}$.
After 4000~s, a single larger downflow event
occurs, which shows up as power at very low frequencies (below the
line marking the cone of influence).

\begin{figure*}
\begin{center}
  \includegraphics[scale=0.6]{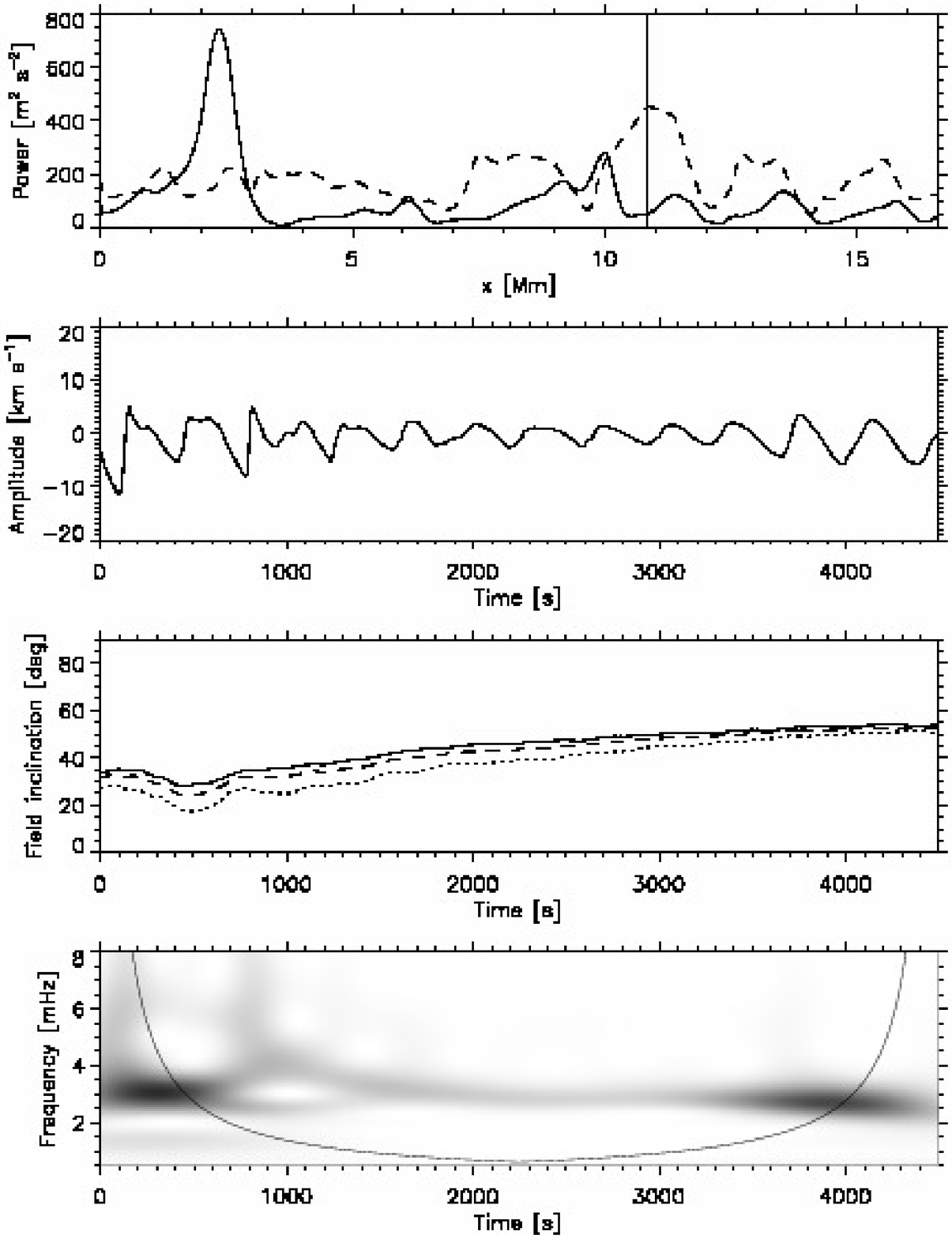}
  \caption{As Figure~\ref{Ewvl073}, but at $x=10.86$~Mm.}
  \label{Ewvl334}
\end{center}
\end{figure*}

The largest peak of the 5-minute power is centered at $x=10.86$~Mm,
shown in Figure~\ref{Ewvl334}. This is directly above a flux concentration
(Figure~\ref{caseEBud}) and there is some periodic velocity signal
throughout the simulation, though the amplitude varies with time. Again,
it is strongest early on, though the peak is mostly within the
region subject to edge effects. Although the inclination
is below 40$^{\circ}$ in the beginning of the simulation, little
coherent signal is visible in the 3-minute band. The 5-minute band
shows power throughout, with a second peak appearing towards the
end (again partially within the cone of influence). The field
inclination slowly increases with time and ends up at 53$^{\circ}$.


\section{Discussion}

\begin{figure*}
  \plottwo{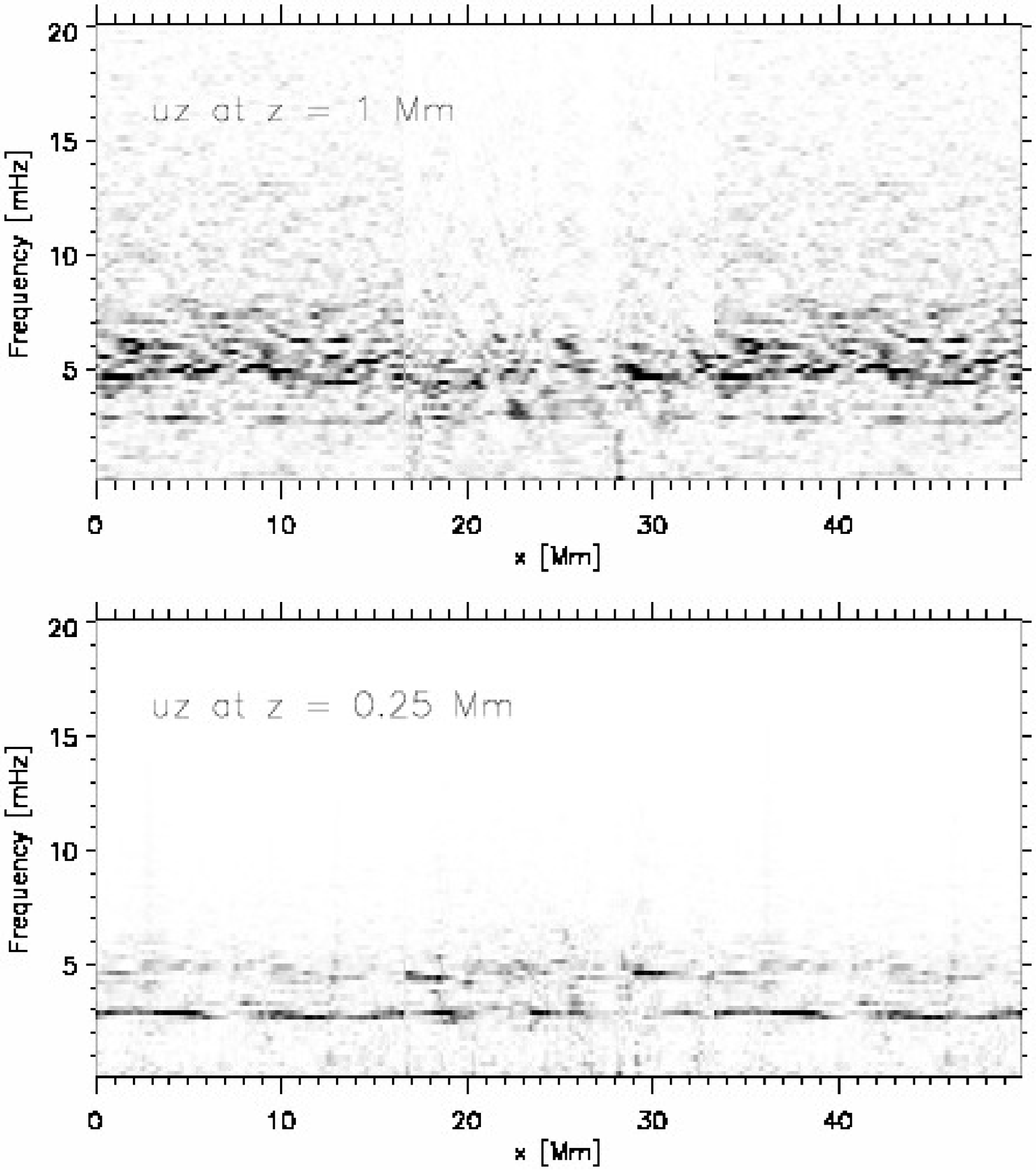}{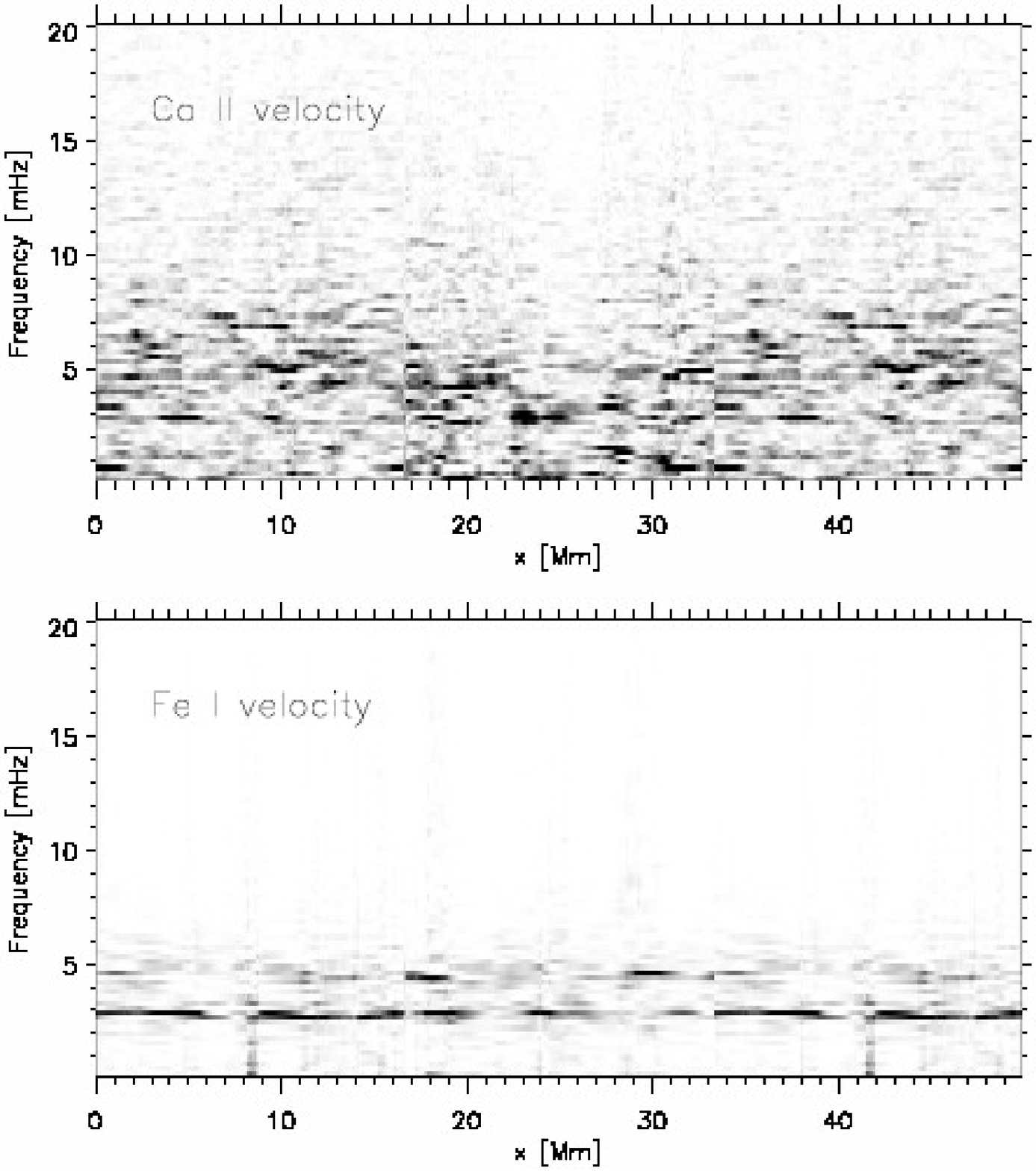}
  \caption{Fourier spectra of the vertical velocity taken from our simulations
           (left panels) and the Doppler velocity derived from synthesized
           spectral lines (right panels). The synthesized spectra have been
           smoothed over 2 arcseconds. In all panels, we show the spectrum
           of case B in the center (between $x=16.5$~Mm and $x=33$~Mm),
           flanked on both sides by the spectrum of the non-magnetic case
           A. This figure can be compared with Figure 4 in
           \citet{Lites+etal1993}.}
   \label{litesrad}
\end{figure*}

The main result from our simulations is that 5-minute waves are able
to propagate through the chromosphere in regions where the magnetic
field is inclined and sufficiently strong. In regions where the field
is vertical or weak, the velocity field is dominated by waves with
periods of around 3 minutes.

As mentioned in the introduction, a model which uses the radiative
relaxation time, rather than the inclination of the magnetic field,
as the mechanism for increasing the cutoff period, has been proposed,
and has been demonstrated to work in numerical simulations
\citep{Khomenko+etal2008}. However, it relies on a simple Newtonian
cooling model for approximating radiative losses.

Our simulations use an advanced and realistic method for computing the
radiative losses
\citep{Gudiksen+etal2011} that includes all the important mechanisms
involved in radiative losses in the photosphere and chromosphere, and as
such it is considerably more realistic than a simple Newtonian cooling
model. While the radiative relaxation time model expects 5-minute
propagation above
all strong small-scale magnetic structures, regardless of the field inclination,
our simulations show 3-minute propagation above the central region of
flux concentrations, and 5-minute propagation in inclined-field regions to
the sides. This is exactly the result predicted by the field inclination model,
but is in conflict with the predictions of the radiative relaxation model.
Our conclusion, based on the simulations presented here, is
therefore that the radiative relaxation model is not effective when
a more realistic energy equation is considered.

While large-scale regions of homogeneous inclined field, like our cases D
and E, represent idealizations of the conditions at the edge of plages,
our case C represents realistic conditions
for an isolated magnetic element or pore. In this model, the magnetic field
is largely vertical in the photosphere and above the transition region, while
the field expansion leads to significant inclination where it is
needed the most,
in the chromosphere. This can then account for the 5-minute propagation, but
we expect 3-minute propagation in the center of the flux concentration. Is
this supported by observations?

The answer is in fact not very clear. \citet{Lites+etal1993} show an example
of a network region which has very little power at frequencies above 4~mHz
over most of its area. There is, however, a pronounced spike of higher-frequency
power located in the center of the network ($x=30$~Mm in their Figure 4).
The authors attribute this to
noise from seeing, which it may very well be; it is visible at frequencies up
to 20~mHz, and a study of the coherence spectra also lends credence to this
explanation. However, the signal at frequencies of 5-6~mHz in this spike 
is much stronger than that at higher frequencies, and indeed is of the same 
magnitude as the (real) signal at those frequencies in the neighboring quiet
regions. It is notable that such a spike, with a spatial extent of 1-2~Mm, is
in fact exactly what we find in our case C. The central spike observed by
\citeauthor{Lites+etal1993} could thus be real.

In order to make more direct comparisons to their results, which
were based on Doppler shifts in observations of the Ca~{\sc ii}~H
line at $3968.49$~\AA{} and the Fe~{\sc i} line at $3966.82$~\AA,
we have computed synthetic spectra in these lines from our simulations
using the non-LTE radiative transfer code MULTI \citep{Carlsson1986}.
The Ca line is treated in full non-LTE, while the Fe line, which is
formed in the wing of the Ca line, is included in the same computation
using additional localized opacity and source function terms based
on LTE. The calculations have been performed column by column, i.e.,
neglecting any radiative interaction in the $x$-direction. The
computed spectra have been smoothed over $2''$, which appears to be
close to the effective resolution of the observations in \citet{Lites+etal1993},
and we have then calculated Doppler shifts from these smoothed
spectra.

It should be noted that Ca~{\sc ii}~H has a very complex line profile,
usually with several emission peaks within the deeper absorption
line. In a dynamic atmosphere, particularly in the presence of strong
shocks, these emission peaks can become very large and are often
asymmetric (so called bright grains), and the line can undergo
central reversal as well. Defining and identifying the line center
of such a line is a non-trivial task.
We have used a method that finds the center of the
region where the intensity is below a certain threshold above the
minimum intensity. This method generally gives acceptable results,
but can not be expected to correspond directly to the velocity at
any one given height in the simulations.

The power spectra of the calculated Doppler velocities are shown in
the two right-hand panels of Figure~\ref{litesrad}; the two left-hand
panels show the power spectra of the vertical velocity taken directly
from our simulation data, at heights corresponding to the approximate
formation heights of the lines. Since the observations of \citet{Lites+etal1993}
covered a larger area than our simulation boxes, we show
a combination of the spectra from two of our simulations in the figure.
In the center of all panels, between $x=16.5$~Mm and $x=33$~Mm, we
show the spectrum of our case B, representing network conditions. On
both sides, from $x=0$~Mm to $x=16.5$~Mm and from $x=33$~Mm to $x=49.5$~Mm,
we show the spectrum of our case A, representing conditions in the
weakly magnetized internetwork. This figure should be compared with
Figure 4 in \citet{Lites+etal1993}.

While our synthesized Ca power spectrum (upper right) is not a perfect match to
the observations of \citeauthor{Lites+etal1993}, there are many
similarities. Of particular note is the difference in the dominant
frequencies between the non-magnetic regions on both sides and
the network region in the center.
The network (case B) is dominated by
lower frequencies, particularly in the area between $x=22$~Mm and
$x=29$~Mm, where the dominant frequencies are 3-$3.5$~mHz. The
internetwork (case A) has a more scattered spectrum, but the dominant
frequencies are mostly between 5 and 7~mHz. Furthermore, the network
has significant power at very low frequencies, below 3~mHz. These
effects are also found by \citeauthor{Lites+etal1993}.

This power at low frequencies is not found in the simulation velocity
at $z=1$~Mm (upper left), and although the velocity spectrum and the Ca spectrum
are similar in many ways, there are also several differences. These
differences are partly due to the smoothing applied to the Ca data,
partly due to the difficulty of defining a meaningful Doppler shift of the
highly complex Ca line profile, and partly due to the fact that the
line is formed over a range of heights rather than at one given height,
and this height range can also vary with both horizontal position
and time.
Using the Ca Doppler velocity as a proxy for
atmospheric velocity on the real Sun can therefore be misleading.

The velocity spectrum of the Fe line (lower right) shows a very good
correspondence with the velocity at $z=0.25$~Mm in the models (lower
left). This line has a much simpler profile and is formed in a region
with few strong shocks affecting local conditions. The dominant
frequency is 3~mHz in both the network and the internetwork, though
there is also some power at 5~mHz in most locations.
\citeauthor{Lites+etal1993} also find that the power in this line
is in the 3-5~mHz band, although most of it is between 3 and 4~mHz.
Like us, they find no significant difference between the spectra
in the internetwork and in the network in this line.

More recent observations related to the question of wave periodicity
have been performed by \citet{Centeno+etal2009}, \citet{deWijn+etal2009},
and \citet{Stangalini+etal2011} \citep[see also][]{Jefferies+etal2006}.
\citeauthor{Centeno+etal2009} used the Tenerife
Infrared Polarimeter of the German Vacuum Tower Telescope at the
Observatorio del Teide, with a seeing-limited spatial resolution
of $1''-1\farcs{}5$. They then found propagating 5-minute waves in the
chromosphere above a facular region. The photospheric magnetic
field as determined from Stokes inversions was within 20$^{\circ}$
of the vertical. \citeauthor{deWijn+etal2009}, using the Solar Optical
Telescope on {\it Hinode} with a resolution of $0\farcs{}16$, found
3-minute signal in the center of a plage region, but more 5-minute
propagation towards the sides in the direction of the expanding field.
\citeauthor{Stangalini+etal2011} used a combination of data from {\it Hinode}
and the IBIS instrument at the Dunn Solar Telescope,
with an estimated average resolution of $0\farcs{}36$.
They found propagating 5-minute waves along the inclined field on
the edges of a pore, and some power in 3-minute oscillations at the
center. They also found both 5-minute and 3-minute propagation, though
with more power in the 5-minute band, in a nearby region
with small magnetic elements where they estimate that the chromospheric
magnetic field is close to vertical, based on a force-free field extrapolation.

The results of our simulations are in agreement with
\citet{deWijn+etal2009}, and with the propagation patterns observed
by \citet{Stangalini+etal2011} around their pore. In the more vertical
magnetic structures observed by \citet{Stangalini+etal2011} and
\citet{Centeno+etal2009}, we would expect more power in the 3-minute band
than in the 5-minute band based on the results of our simulations,
if indeed the field is mainly vertical and the flux tubes do not move
around very much. There are, however, several
possible mechanisms that could explain the 5-minute dominance and
resolve this apparent difference.

For one, all flux tubes naturally expand with height, and this expansion
creates a region of inclined field (as illustrated, on a
large scale, by our case C). Thus, even if the field is close to vertical
in the photosphere, there will be regions between the photosphere
and the chromosphere where the field at the edges of the flux tubes
is inclined, and the long-period waves can propagate there.
\citet{Centeno+etal2009} do study coherence spectra to look
for signs of a possible horizontal shift in the signal as a result
of propagation along inclined fields, and find good coherence
between the photospheric
and chromospheric signal, but our results show that the field does
not need to be inclined throughout the photosphere and chromosphere
in order to enable 5-minute wave propagation. A few hundred km along
the edges of an expanding flux tube may be enough, and any horizontal
shift could then be less than one resolution element ($1''-1\farcs{}5$
in their data). In the observations of \citet{Stangalini+etal2011},
the photospheric field is not uniformly vertical, and a force-free
extrapolation is not a very good approximation in the chromosphere.
Furthermore, although there is more power in the 5-minute signal,
they also find significant signal at periods around
3 minutes in the region with smaller magnetic elements. We believe
that our model, where field inclination is the dominant mechanism
for allowing long-period wave propagation, is compatible with these findings.

Flux tube movement and limited resolution may also be partially
responsible for the relative dominance of 5-minute power in the
observations of \citet{Centeno+etal2009}. In case B, we found that
the strongest 3-minute power appeared above flux tubes that
undergo little horizontal motion, while 5-minute power was found
in inclined field regions at the edges of flux tubes. If the
flux tubes move around, both the 5-minute and 3-minute power
will be spread out and one would not see a clear distinction
between the (average) flux tube center and the sides in a
Fourier analysis. The flux tube center also covers
a rather small area at any given time,
and this makes the related 3-minute waves difficult to
observe in low-resolution data. In the higher-resolution
data of \citet{Stangalini+etal2011}, regions of 3-minute
propagation are found, and this could possibly be because
the flux tubes at these locations move around less.
We would encourage observers to
look for differences in the periodicity of oscillations at
the center and edges of flux tubes in future high-resolution datasets.

A third possibility is that heating may play a role. The
temperature structure of the chromosphere is in general not
well known. In particular, the real Sun may have more magnetic
heating of the upper chromosphere than our 2D models. Higher
temperature would reduce the cutoff frequency and allow 5-minute
waves to propagate more easily.

Yet another suggestion, as mentioned by \citet{deWijn+etal2009},
is that the field may be twisted. The waves could then travel
along field lines that are everywhere inclined with respect to the
local vertical, but without significant horizontal displacement.
Such field twist is a 3D effect and can not be tested in our 2D simulations,
but should be considered in later work.

A different point, that we have already mentioned in our analysis,
is also worth making: although a Fourier analysis can be a powerful
tool, it has some important limitations. In order to achieve
sufficient spectral resolution, one usually needs time series on the
order of one hour. The solar atmosphere, however, is dynamic on
timescales of minutes. Atmospheric conditions
can and do change, and the Fourier transform is not well suited
for picking up such changes. At least in areas where the general
signal is weak, non-recurring events can end up dominating the
power spectrum (see Figure~\ref{bampang456} for an example; there
are several other examples in the dataset). In such
cases, the Fourier analysis does not say all that much about
the general conditions at that location. This can be particularly
dangerous because the eye is naturally drawn to peaks in the
power spectrum. Also, there can be times when local conditions are notably
different from the time average, and these can be correlated with
changes in the signal (e.g., Figure~\ref{vertpwrudang185}).
A wavelet analysis, which takes time variations into account,
or at least a careful examination of the time series,
is essential for identifying the actual processes going on
in the atmosphere.

\section{Jet formation}

So far we have been looking at the propagation of waves through
the chromosphere. But one of the reasons why we are interested in
this propagation is the effect these waves have once they reach
the transition region.

Jets of chromospheric material go by many names, such as spicules, mottles,
fibrils, straws, macrospicules, or surges. There is no widespread agreement on
whether all or some of these represent different aspects of the
same underlying phenomenon, or on the driving mechanism or mechanisms
behind them. However, there is mounting evidence that at least
dynamic fibrils and some mottles appear to be driven by waves
coming from below \citep{Hansteen+etal2006,DePontieu+etal2007,
Rouppe+etal2007,Heggland+etal2007}. Such wave-driven jets also
appear frequently in our simulations here, and we will look into
their properties and make a statistical comparison between jets
in case B (mostly vertical field) and case D (field inclined 45$^{\circ}$).

\begin{figure*}
\begin{center}
  \includegraphics[scale=0.8]{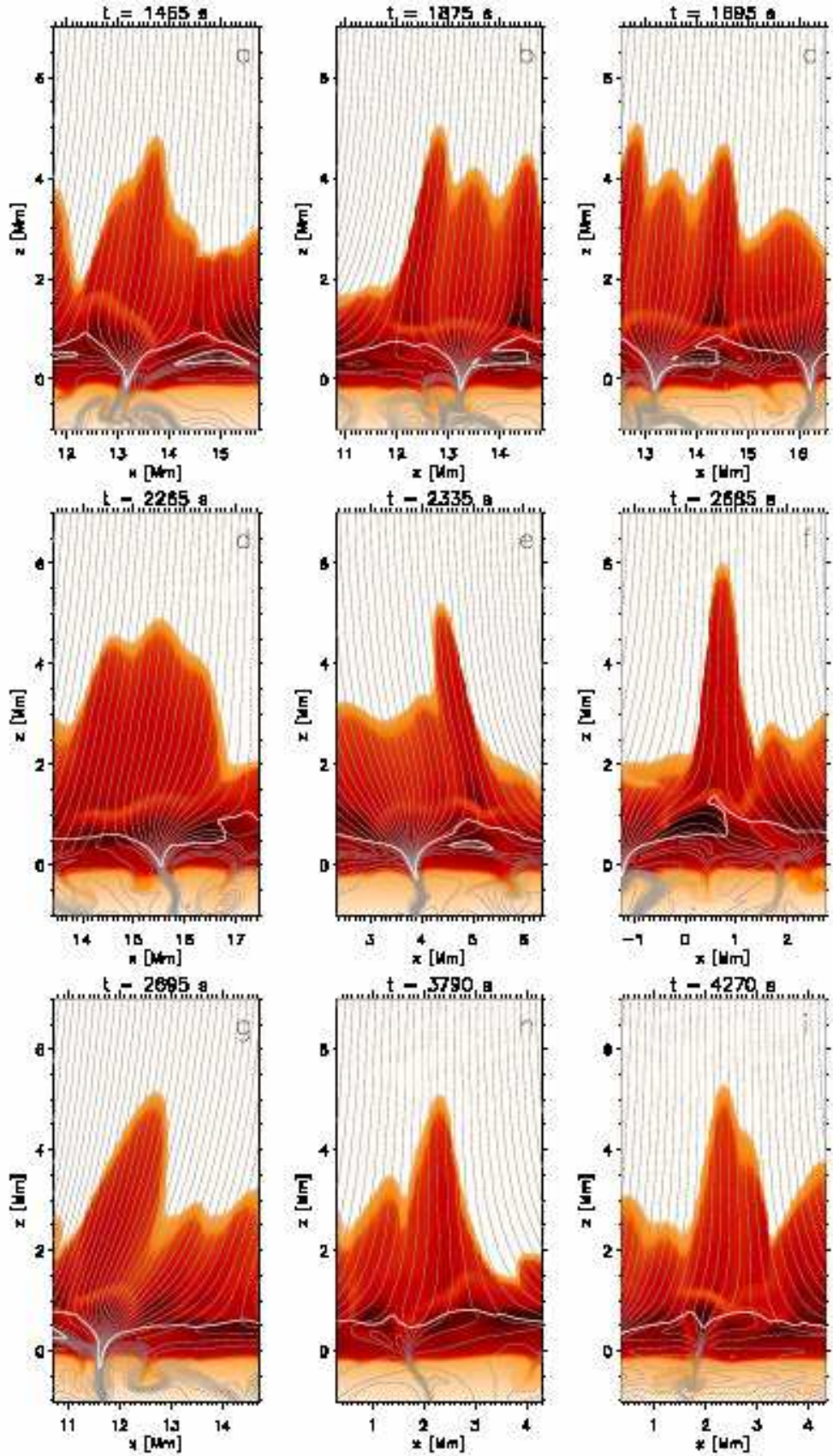}
  \caption{Temperature plot showing a selection of the tallest jets
           in case B. The
           gray lines are magnetic field lines, while the white line
           is at the height where $\beta=1$.}
  \label{vertspic9}
\end{center}
\end{figure*}

\begin{figure*}
\begin{center}
  \includegraphics[scale=0.8]{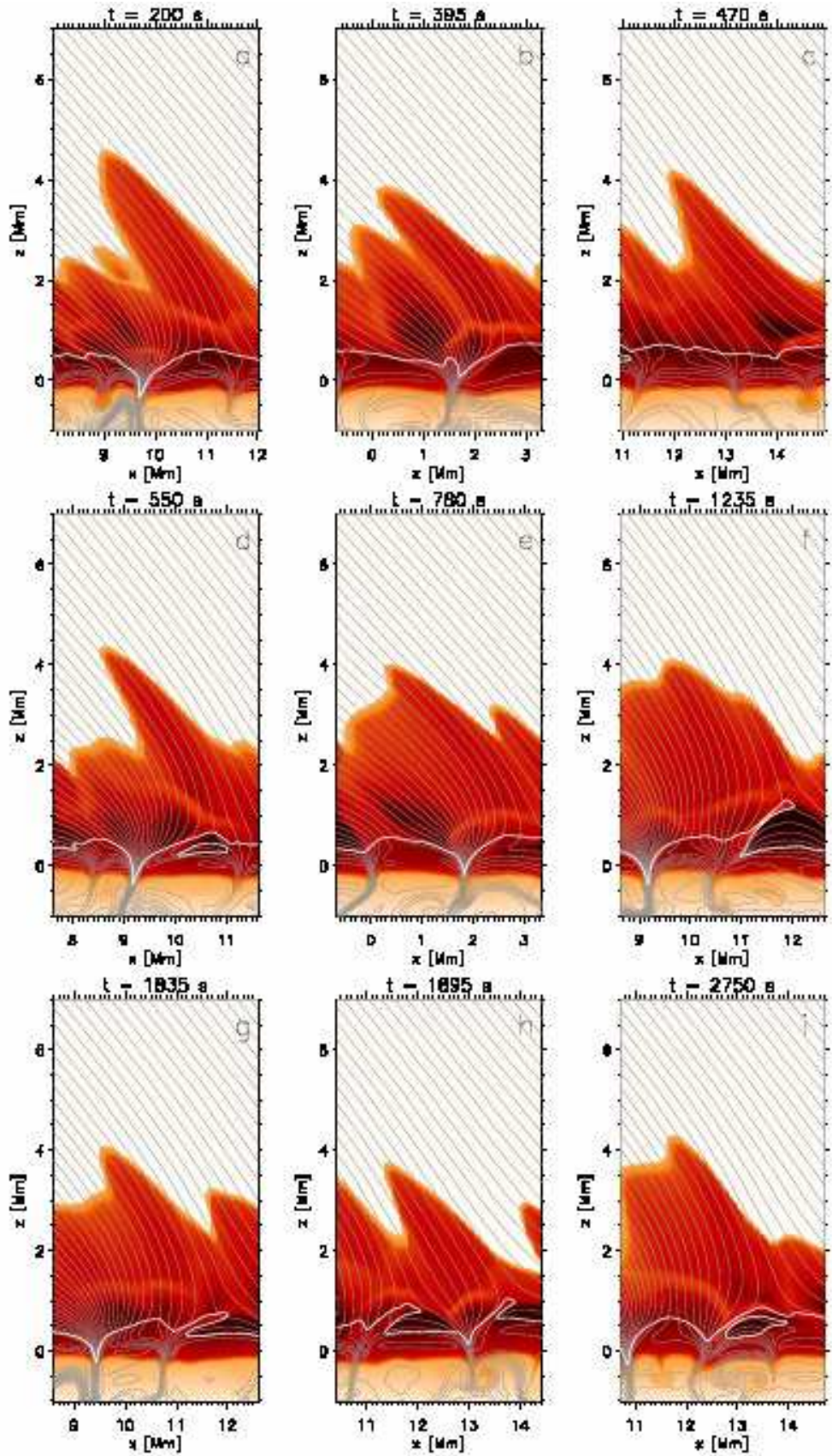}
  \caption{Temperature plot showing a selection of the tallest jets in case D.}
  \label{45degspic9}
\end{center}
\end{figure*}

A large number of jets are formed in the simulations, and we
have used semi-automatic routines to identify and measure them.
We identify the jets as local maxima (in both space and time) of
the height where the temperature is 40~000~K, corresponding to
the lower transition region. We have included snapshots of the tallest
jets in the two simulations in Figures~\ref{vertspic9} (case B)
and \ref{45degspic9} (case D). We see that they can be quite tall,
reaching heights of up to 6~Mm, and that their axes follow the
magnetic field, as we would expect.

\begin{figure*}
\begin{center}
  \includegraphics[scale=0.6]{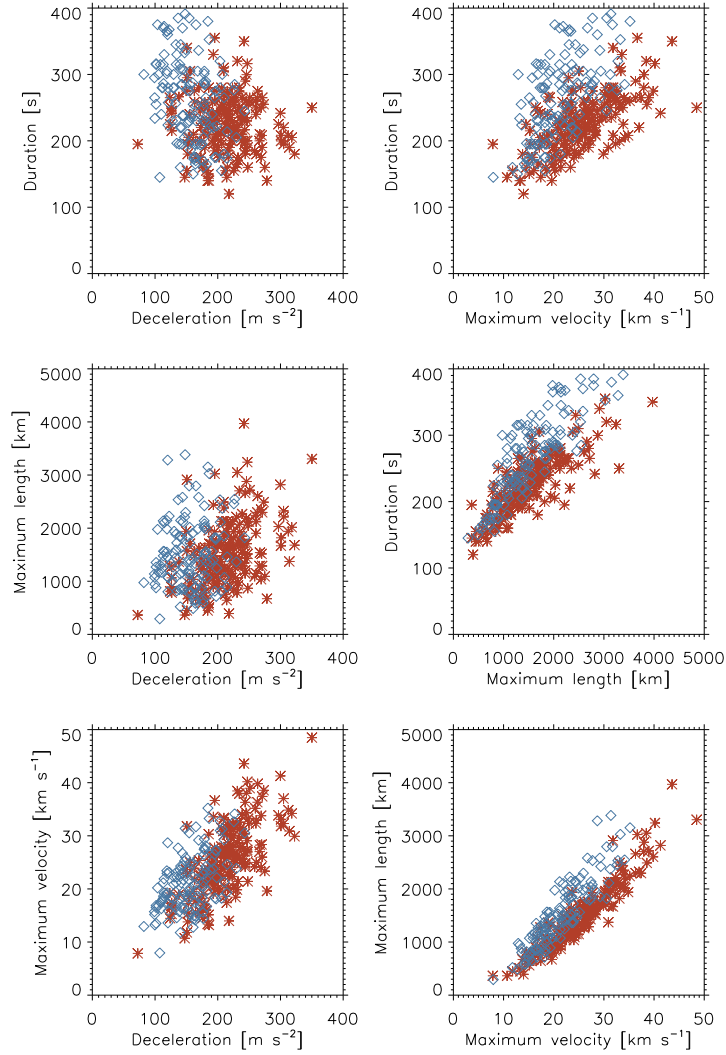}
  \caption{Scatter plots showing correlations between different jet
           properties in case B (vertical field, shown as red asterisks)
           and case D (inclined field, shown as blue diamonds). Most
           quantities exhibit linear correlations; a particularly nice
           one is the one between maximum velocity and maximum length
           (bottom right). There is however no strong correlation between
           the deceleration and either the duration (top left) or the
           maximum length (center left).}
  \label{spiccorr}
\end{center}
\end{figure*}

\begin{figure*}
\begin{center}
  \plotone{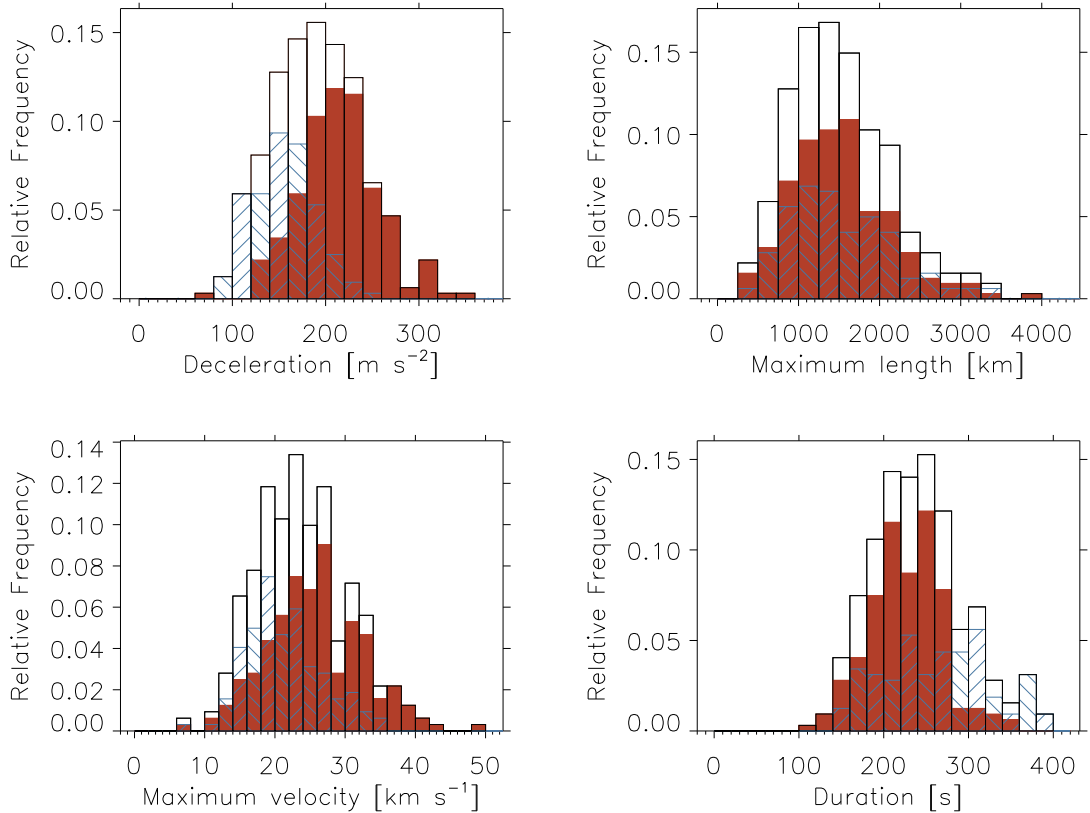}
  \caption{Histograms showing the distributions of deceleration (top left),
           maximum length (top right), maximum velocity (bottom left),
           and duration (bottom right). Case B (vertical field) is shown
           as red solid columns, case D (inclined field) is shown as blue
           hatched columns, and the sum is shown as black open columns.}
  \label{spichist}
\end{center}
\end{figure*}

\begin{figure}
  \includegraphics[scale=0.7]{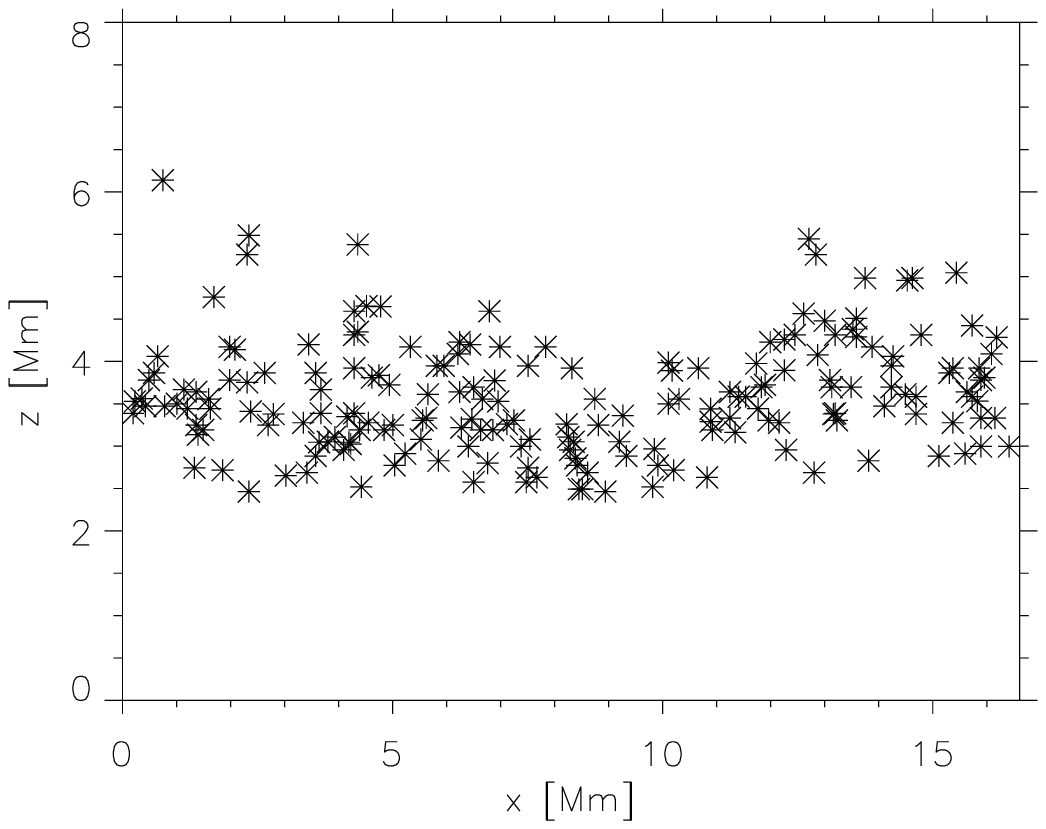}
  \caption{Scatter plot of jet heights versus
           horizontal position in case B.}
  \label{vertjetzvsx}
\end{figure}

In order to analyze the properties of the jets, we want to look
at their time evolution. A rising and falling jet will generally
have a parabolic shape in a plot of distance vs. time
\citep[e.g.,][]{Hansteen+etal2006}, but for the most accurate
results, the measurements should be taken close to the central
axis of the jet. The jet peaks found previously are used as
starting points. In case D, where the magnetic field is inclined,
we track the magnetic field line that passes through the jet
peak, and measure the variables along this (moving) line through
the jet's lifetime. In case B, the magnetic field and the jet
axes are generally close to vertical, and we use a simplified
procedure taking vertical cuts through the locations of the jet
peaks. Experiments using the more involved field-tracking procedure
in case B show that the inaccuracies resulting from the simplified
treatment are small.

Using a method similar to the one of \citet{Hansteen+etal2006} and
\citet{DePontieu+etal2007}, working on observations, \citet{Heggland+etal2007},
working on 1D simulations, and \citet{Martinez+etal2009a}, working on
3D simulations, we
produce distance-time diagrams of the temperature and fit the position
of the transition region with parabolas.
These fits are then used to calculate properties
such as the maximum velocity, deceleration, maximum length,
and duration of each jet. Jets that do not have a parabolic shape
or that represent only minor disturbances are discarded from the analysis.

Note that, as jets sometimes occur in quick succession in the same
locations, occasionally they can overlap. This can lead
to situations where the descending phase of one jet is cut short
by the next jet appearing. This will mainly affect the calculated
lifetime of the jet, though it can to a lesser extent also impact
the maximum measured length and velocity. The calculated deceleration depends
only on the overall shape of the parabola and is largely unaffected.
Overall, this does not happen frequently enough to significantly
change the statistics, though it can lead to some outliers,
particularly with respect to the durations.

In total, we have identified 192 parabolic jets in case B, and 129
in case D. Their properties and various correlations are shown in
Figure~\ref{spiccorr}, which is in the same format as Figure~1 in
\citet{Martinez+etal2009a} for ease of comparison. They should also
be compared with Figures~12 and 13 in \citet{DePontieu+etal2007}.

Overall, the results of the different simulations and the observations
are in good agreement.
As we see, there are clear linear correlations between several
of the jet properties, in particular between maximum velocity and
maximum length (bottom right), deceleration and maximum velocity
(bottom left), and maximum length and duration (center right). The
same correlations are found in observations by \citet{DePontieu+etal2007},
and in simulations by \citet{Heggland+etal2007}, who only plot deceleration
vs. maximum
velocity, and \citet{Martinez+etal2009a}, who find a weaker
correlation between length and duration. In addition, we find a
correlation between the maximum velocity and the duration (top right)
in our simulations, whereas \citeauthor{DePontieu+etal2007} and
\citeauthor{Martinez+etal2009a} find little evidence of a correlation
here. Conversely, they both find a weak anticorrelation between
deceleration and duration (top left). In our simulations, there is
a quite weak anticorrelation in the distribution for case D (blue
diamonds), whereas case B (red asterisks) shows little evidence
of a correlation. The two distributions do however cluster in different
regions, with the longer-duration jets in case D having lower
decelerations than the shorter-lived jets in case B. This
regional difference between the distributions was also found
in the observations.
Neither we, \citeauthor{Martinez+etal2009a}, nor \citeauthor{DePontieu+etal2007}
find any strong correlation between the deceleration
and the maximum length of the jets (center left).

As noted above, the properties of the jets
formed in case B and case D can be somewhat different --- for example,
at a given maximum velocity, the jets in case D (inclined field)
tend to be longer (bottom right panel) and have a longer duration
(top right panel). In order to further investigate these differences,
we have made histograms showing the distributions of the jet
properties in the two cases. The results are plotted in
Figure~\ref{spichist}, which shows the jets in case B as solid red
columns, case D as hatched blue columns, and the sum as black outlines.
The distributions are notably different with respect to deceleration
(upper left panel) and duration (lower right), with the inclined
jets lasting longer and having lower decelerations. The distributions
of maximum lengths (upper right) are broadly similar, but the inclined
jets have somewhat lower maximum velocities (lower left).
\citet{DePontieu+etal2007} found similar regional differences in
their observations, with jets in dense plage regions (referred to
as region 2 in their paper) having higher decelerations and
shorter lifetimes than jets in adjacent regions with weaker
and more inclined field (referred to as region 1). They suggested
that the differences were due to the field inclination, which is supported
by our results. The results are also consistent with what we have
found out about wave propagation in the different models: long-period
waves are more prevalent in inclined field regions, and these waves
then produce longer-lived jets with lower decelerations.

In addition to differences between the cases, there are also some
regional differences within each case. We see in Figure~\ref{vertspic9}
that a number of the tallest spicules in case B are located in regions
where field lines from different flux concentrations meet (ref.
Figure~\ref{vertBud}). This is
particularly noticeable in the cases of jets $e$ and $f$, where the
central axis is right along the interface region. Jets $b$, $c$, $h$, and $i$
are also located close to the interface, while jets $a$ and $g$ are
above the center of their connected flux concentrations.

In Figure~\ref{vertjetzvsx}, we have plotted the maximum heights
reached by the jets as a function of their horizontal location.
Notably, the jets that form above the central, wide flux concentration
($x=8-11$~Mm) reach lower heights than the ones forming in other locations,
while the tallest jets form in interface regions between flux concentrations
at $x=13-15$~Mm and $x=0-4$~Mm. There are at least two possible
explanations why this should be so. One is that you can get 
constructive interference between the waves coming from different
flux concentrations, leading to stronger shocks and longer jets.
For example, jet $e$ in Figure~\ref{vertspic9} can be traced to such
an event happening at $x=5$~Mm, $t=2150$~s (cf. Figure~\ref{vertBud},
which shows the velocity field). Another
possible explanation is that when several flux concentrations (of
the same polarity) are located near each other, the field lines
connected with each cannot spread out over such a large area, and
the waves coming from below will be more concentrated and propagate more
vertically. Conversely, above the large central flux concentration,
the wave energy is spread over a much larger area as much of it follows
the widening field. This then leads to wider but less powerful jets.

\section{Summary}

We have presented several simulations investigating the periodicity
of waves propagating through the chromosphere. We find that waves
with periods of around 5 minutes can propagate in regions where the
magnetic field is strong and inclined, including at the edges
of flux tubes. In regions where the magnetic
field is weak or vertical, we find primarily 3-minute waves; this
also applies above vertical flux tubes and in the center of strong
expanding flux tubes. These results indicate that field inclination
is critical to the propagation of long-period waves. Where the flux
tubes undergo significant horizontal motion, both the 5-minute
and the 3-minute power is spread out and the distinction is not
as clearly visible. Since we
have included an advanced treatment of radiative losses in our simulations
and find 3-minute propagation above vertical field regions, we
conclude that variation in the radiative relaxation time is not
an effective mechanism for increasing the cutoff period. Our
simulations are in agreement with the results of recent high-resolution
{\it Hinode} observations.

We have also studied the jets produced by these waves once they
reach the transition region. We find systematic differences between
the jets produced in a model with mostly vertical field, and in
a model with mostly inclined field. The results are in agreement
with observations of dynamic fibrils.

It is also important to point out, for the purpose of
future analyses of wave propagation and periodicity, whether from
simulations or observations, that
the Fourier analysis can be misleading and hide important information
about the state of the medium. This is because the solar atmosphere
is dynamic and changes on timescales much shorter than one hour,
which is often the minimum timescale needed to achieve sufficient
spectral resolution in a Fourier analysis.
A wavelet analysis takes variations
in time into account and can be an invaluable
tool for figuring out what processes are important for the dynamics.

\acknowledgements
This work was supported by the Research Council of Norway through grant
159483/V30. Bart De Pontieu was supported by NASA grants NNX08AL22G and
NNX08BA99G.



\end{document}